\documentclass[twocolumn,pre,amsmath,amssymb]{revtex4-1}

\usepackage{graphicx}
\usepackage{color}
\usepackage{dcolumn}
\usepackage{bm}
\usepackage{float}
\usepackage[table]{xcolor}
\usepackage[utf8]{inputenc}

\newcommand{\cK}{\ensuremath{\mathcal{K}}}
\newcommand{\cG}{\ensuremath{\mathcal{G}}}
\newcommand{\cH}{\ensuremath{\mathcal{H}}}

\newcommand{\cS}{\ensuremath{\mathcal{S}}}

\newcommand{\bx}{\ensuremath{\boldsymbol{x}}}
\newcommand{\by}{\ensuremath{\boldsymbol{y}}}

\newcommand{\bth}{\ensuremath{\boldsymbol{\theta}}}

\newcommand{\bqa}{\ensuremath{\boldsymbol{q}^{(1)}}}
\newcommand{\bqb}{\ensuremath{\boldsymbol{q}^{(2)}}}

\newcommand{\bbqb}{\ensuremath{\bar{\boldsymbol{q}}^{(2)}}}

\newcommand{\bLg}{\ensuremath{\boldsymbol{L}}^{\text{rw}}_{\mathcal{G}}}
\newcommand{\bLh}{\ensuremath{\boldsymbol{L}}^{\text{rw}}_{\mathcal{H}}}

\newcommand{\bI}{\ensuremath{\boldsymbol{I}}}
\newcommand{\bW}{\ensuremath{\boldsymbol{W}}}

\newcommand{\bE}{\ensuremath{\boldsymbol{E}}}
\newcommand{\bDelta}{\ensuremath{\boldsymbol{\Delta}}}
\newcommand{\bu}{\ensuremath{\boldsymbol{u}}}
\newcommand{\bpg}{\ensuremath{\boldsymbol{p}_{\mathcal{G}}}}
\newcommand{\bph}{\ensuremath{\boldsymbol{p}_{\mathcal{H}}}}

\newcommand{\dgG}{\ensuremath{D_{(l)}^{(3)}(\cG)}}
\newcommand{\dgH}{\ensuremath{D_{(l)}^{(3)}(\cH)}}
\newcommand{\dgX}{\ensuremath{\mathcal{S}_{(l)}^{(2)}(X)}}
\newcommand{\dgY}{\ensuremath{\mathcal{S}_{(l)}^{(2)}(Y)}}
\newcommand{\dgtG}{\ensuremath{D_{(l),\tau}^{(2)}(\cG)}}
\newcommand{\dgtH}{\ensuremath{D_{(l),\tau}^{(2)}(\cH)}}

\newcommand{\dHau}{\ensuremath{d_\textup{H}}}
\newcommand{\dBot}{\ensuremath{d^{(2)}_\textup{B}}}

\newcommand{\dBBxi}{\ensuremath{d^{(3)}_{\textup{B},\xi}}}

\newcommand{\bPGtau}{\ensuremath{P_{\cG}(\tau)}}
\newcommand{\bPHtau}{\ensuremath{P_{\cH}(\tau)}}

\newcommand{\vrip}{\ensuremath{V_\textup{R}}}
\newcommand{\tmax}{\ensuremath{\tau_\textup{max}}}
\newcommand{\kfdr}{\ensuremath{\kappa}}

\newcommand{\rst}{\cellcolor[RGB]{240, 153, 149}}
\newcommand{\rnd}{\cellcolor[RGB]{249, 207, 205}}

\newcommand{\st}{\shortstack}

\usepackage{hyperref}
\hypersetup{
    colorlinks = true,
    linkcolor = {blue},
    citecolor = {blue},
    urlcolor = {blue},
}

\begin{document}
\title{Scale-variant topological information for characterizing the structure of complex networks}

\author{Quoc Hoan Tran}
\email{zoro@biom.t.u-tokyo.ac.jp}
\affiliation{
	Department of Information and Communication Engineering, Graduate School of Information Science and Technology, The University of Tokyo, Tokyo 113-8656, Japan
}

\author{Van Tuan Vo}
\email{tuan@biom.t.u-tokyo.ac.jp}
\affiliation{
	Department of Information and Communication Engineering, Graduate School of Information Science and Technology, The University of Tokyo, Tokyo 113-8656, Japan
}

\author{Yoshihiko Hasegawa}
\email{hasegawa@biom.t.u-tokyo.ac.jp}
\affiliation{
	Department of Information and Communication Engineering, Graduate School of Information Science and Technology, The University of Tokyo, Tokyo 113-8656, Japan
}

\date{\today}

\begin{abstract}
The structure of real-world networks is usually difficult to characterize owing to the variation of topological scales, 
the nondyadic complex interactions, and the fluctuations in the network. 
We aim to address these problems by introducing a general framework using a method based on topological data analysis.
By considering the diffusion process at a single specified timescale in a network, 
we map the network nodes to a finite set of points that contains the topological information of the network at a single scale. 
Subsequently, we study the shape of these point sets over variable timescales that provide scale-variant topological information,
to understand the varying topological scales and the complex interactions in the network.
We conduct experiments on synthetic and real-world data to demonstrate the effectiveness of the proposed framework 
in identifying network models, classifying real-world networks, and detecting transition points in time-evolving networks.
Overall, our study presents a unified analysis that can be applied to 
more complex network structures, as in the case of multilayer and multiplex networks.
\end{abstract}

\pacs{Valid PACS appear here}

\maketitle

\section{Introduction}

\begin{figure}
		\includegraphics[width=8.5cm]{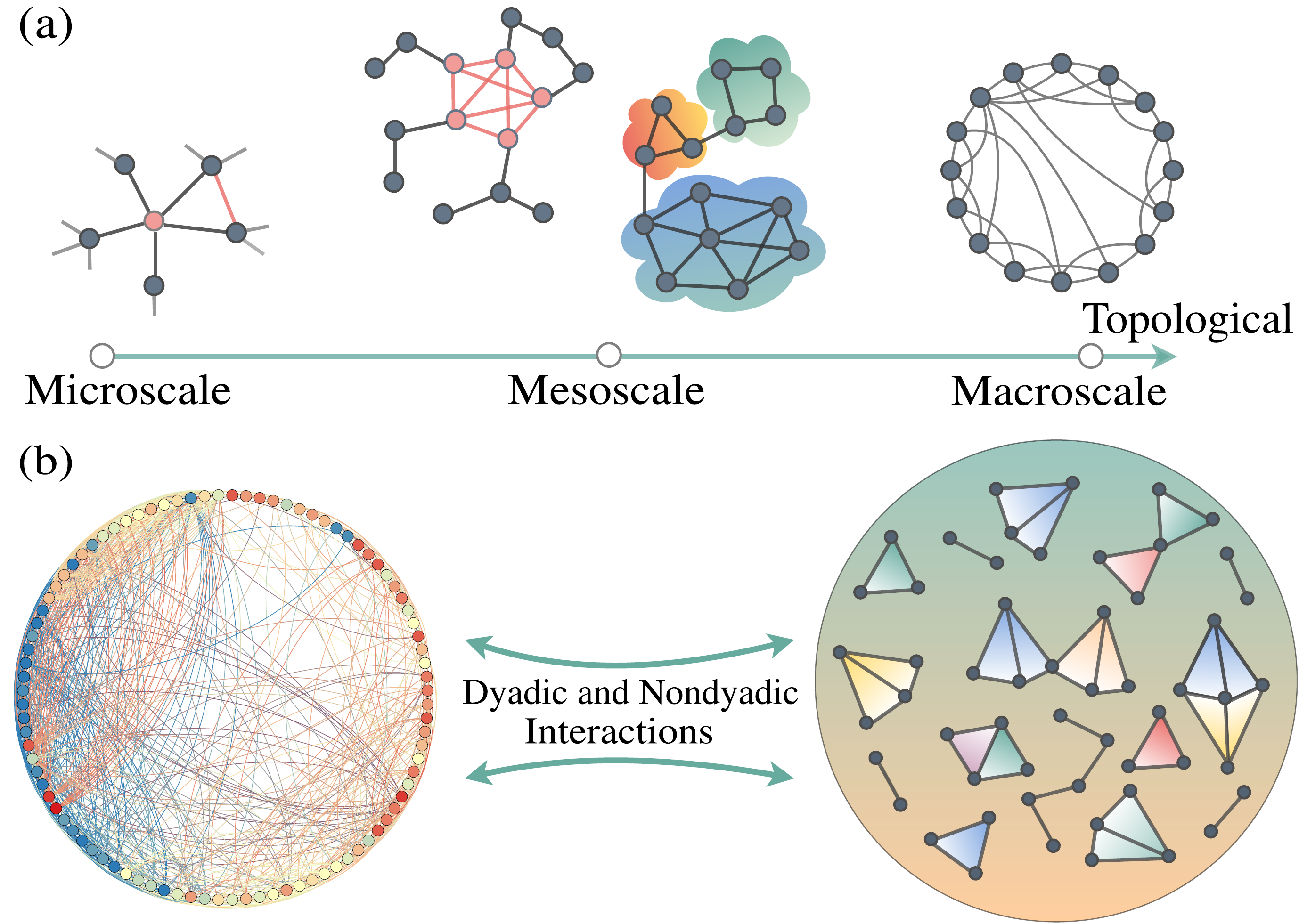}
		\protect\caption{Various topological scales and interactions between multiple elements in a complex network.
		(a) Complex networks can be analyzed at various topological scales ranging from individual nodes (microscale) 
		to the whole network (macroscale).
		In between the two scales, there is a mesoscale, where we can observe patterns of collectives, cores, and peripheries.
		(b) Complex network is a representation of a complex system having dyadic and nondyadic interactions between its elements.
		The interactions can be represented as simplices such as segments (for dyadic interactions), 
		filled triangles, or filled tetrahedrons (for nondyadic interactions involving three or four elements), and so on.
		\label{fig:multiscale}}
\end{figure}

Characterizing the structure of complex networks is the most fundamental challenge in deciphering network dynamics.
The anatomy of a network is quite relevant to phenomena occurring in networks,
such as the spread of information, epidemic disease, or robustness under attack.
Moreover, it has attracted considerable research interest given the numerous applications
including controlling and predicting patterns of dynamics in networks~\cite{taylor:2015:topological,zanudo:2017:structure,santolini:2018:pertubation},
evaluating the structural and functional similarities of biological networks~\cite{sun:2014:predicting,calderone:2016:desease,schieber:2017:quantification},
and detecting transition points in time-evolving 
networks~\cite{carpi:2012:evol,barnett:2016:change,bao:2018:core}.
In a technical sense, the structure of real-world networks is inherently difficult to characterize,
firstly, because these networks have complex patterns that can reflect various topological scales ranging from microscale (individual nodes) to mesocale (community, cores, and peripheries), to macroscale (the whole network)~\cite{ahn:2010:link,betzel:2017:multi,boulos:2017:multi}~[Fig.~\ref{fig:multiscale}(a)].
For demonstrating these patterns, the conventional statistical 
measures~\cite{costa:2007:characterization,newman:2010:networks} and  methods~\cite{arenas:2006:prl,sales:2007:hierarchical,lanci:2009:hierarchical,ahn:2010:link,tremblay:2014:graph} 
are limited when representing the varying topological scales.
Secondly, real-world networks represent complex systems that have dyadic and nondyadic interactions~\cite{marvel:2011:continuous,van:2013:mathematical,reimann:2017:clique}~[Fig.~\ref{fig:multiscale}(b)].
Majority of the current methods used for characterizing complex networks 
focus only on the dyadic interactions,
such as detecting the existence of pairwise edges or paths connected by successive edges.
Thirdly, real-world networks often suffer from fluctuations caused by external factors~\cite{menezes:2004:fluctuation}.
Consequently, the quest for unifying the principles underlying 
the topology of networks emerges only in simple, idealized models~\cite{barabasi:2016:network,broido:2019:scale}.

Herein, we propose a general framework for characterizing the structure of complex networks,
mainly based on the topological data analysis of a 
diffusion process viewed at variable timescales.
We consider a diffusion process in which a random walker moves randomly between nodes in continuous time 
at the transition rate proportional to the edge weights.
The interaction between the nodes via the diffusion process 
can reflect the structure of the network at different topological scales.
For example, a microscale structure is revealed with a small diffusion timescale $\tau$.
Increasing $\tau$ will increase the ranges of interactions to 
reflect the mesoscale decomposition of the network, 
until the macroscale structure is finally captured.
By considering the diffusion process at a single specified timescale $\tau$,
we can map the network nodes to a finite set of points known as 
a \textit{point cloud} in a high dimensional space.
In the point cloud, a group of close points represents 
the unit of interacted nodes in the diffusion process.
The shape of this point cloud contains the topological information 
of the network at a single topological scale.

Based on a topological data analysis method that provides insight
into the ``shape'' of data~\cite{carlsson:2009:topology},
we build a geometrical model that is primarily a collection of geometrical shapes
to reveal the underlying structure of the point cloud.
In this geometrical model, two points in the point cloud are connected if their distance is less than or equal to a given threshold.
If the threshold is considerably small, only points appear in the geometrical model,
and no connections are created between points.
As the threshold is gradually increased, more pairwise connections are created, 
and geometrical shapes as line segments, triangles, tetrahedrons, and so on, are added to the geometrical model.
In the case where the threshold becomes considerably large, all pairs of points in the point cloud will be connected,
and only a giant overlapped geometrical shape remains in the space.
To obtain information regarding the ``shape'' of the point cloud,
we focus on the changes of topological structures, 
such as the merging of connected components, 
and the emergence and disappearance of loops in the geometrical model as the threshold is increased.
Therefore, at each timescale $\tau$, we construct the topological features to 
monitor the emergence and disappearance of the topological structures.
We can consider such features as a representation for the network at a single topological scale ($\tau$-scale).
Further, we extend these features by considering the timescale $\tau$ as a variable parameter
instead of a single fixed value.
The extended features, referred to as scale-variant topological features, 
can reflect the varying topological scales in the complex network.

The scale-variant topological features are proven to be robust
under perturbation applied to the network, and thus,
can serve as discriminative features for characterizing the networks.
We input these features in the kernel technique in machine learning algorithms to apply to
statistical-learning tasks, such as classification and transition points detection.
We show that the proposed framework can characterize the parameters that are used to generate the networks through an analysis of several network models.
Furthermore, we can classify both synthetic and real-world networks 
with more effective results when compared with other conventional approaches.
We further apply the proposed framework to detect the transition points with respect to the topological structure
in the time-evolving gene regulatory networks of \textit{Drosophila melanogaster}.
Interestingly, these transition points agree well with the transition points relative to the dynamics 
obtained from the experimental results on the profiling.

\section{Method}
\subsection{Scale-variant topological features}
Let $\cG$ be an undirected weighted network with $N$ nodes, $v_1,\ldots,v_N$,
and assume that there is a single random walker moving randomly between the nodes in continuous time.
When the walker is located at $v_i$, we assume the walker to move to the neighboring node $v_j$ 
at a transition rate $w_{ij}/W_i$, where $w_{ij}\geq 0$ represents 
the weight of the edge from $v_i$ to $v_j$ ($i,j\in \{ 1,2,\ldots,N \}$) and $W_i=\sum_{j=1}^Nw_{ij}$.
Herein, if there is no edge between $v_i$ and $v_j$, then $w_{ij}=0$.
Now, let $p_{\cG, k}(\tau|i)$ denote the probability 
of a random walker on $v_k$ at time $\tau$ that starts from $v_i$.
The probability distribution vector, $\bpg(\tau|i)=\left[p_{\cG, 1}(\tau|i),\ldots,p_{\cG, N}(\tau|i)\right]$, 
is given based on
the solution of the Kolmogorov forward equation~\cite{domenico:2017:diffusion}:
\begin{align}\label{eqn:master}
    \dfrac{d\bpg(\tau|i)}{d\tau}=-\bpg(\tau|i)\bLg.
\end{align}
Here, $\bLg$ is the random walk Laplacian whose components $l_{ij}$ ($i,j\in \{ 1,2,\ldots,N \}$) are given by,
\begin{align}
 l_{ij} = 
    \begin{cases} 
      1  &\textrm{ if $i=j$ and $W_i \neq 0$}\\
      -w_{ij}/W_i  &\textrm{ if $i\neq j$ and $v_i$ is adjacent to $v_j$}\\
      0  &\textrm{ otherwise.}\\
    \end{cases}
\end{align}
The solution for Eq.~\eqref{eqn:master} is $\bpg(\tau|i)=\bu_i\exp(-\tau\bLg)$, 
where $\bu_i=\left[0,\ldots,0,1,0,\ldots,0\right]$ with its $i$-th 
element being equal to 1; the others are equal to 0 ($i\in \{ 1,2,\ldots,N \}$).

\begin{figure*}
		\includegraphics[width=16cm]{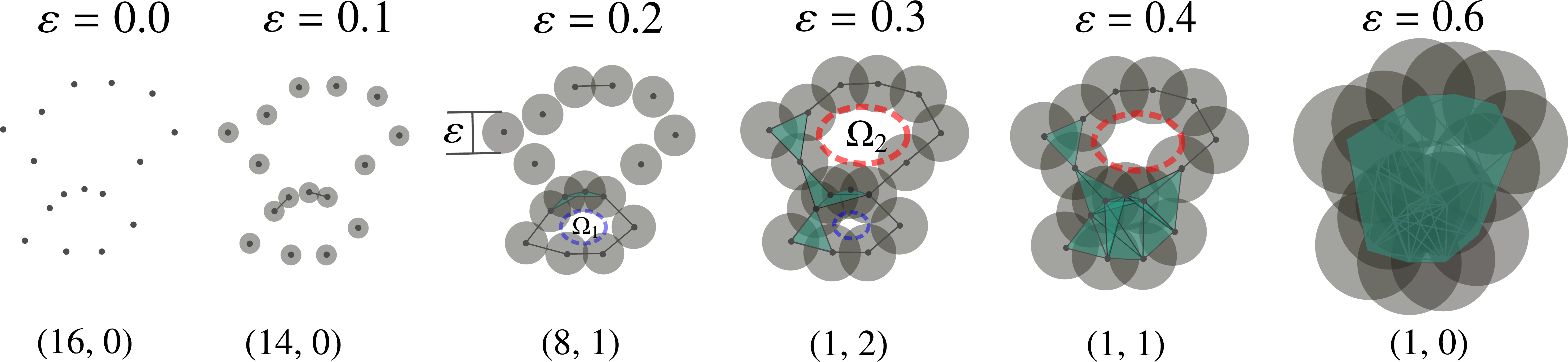}
		\protect\caption{
		An exemplary of Vietoris--Rips filtration constructed from a point cloud.
		A union of balls of radius $\varepsilon/2$ centered at each point is considered.
		Topological structure changes are tracked, such as 
		the merging of connected components or clusters,
		and the emergence and disappearance of loops or circular holes present in the space
		with increasing $\varepsilon$ from \{0.0, 0.1, 0.2, 0.3, 0.4, 0.6\}.
		For instance, the blue loop $\Omega_1$ appears at $\varepsilon=0.2$ then disappears at $\varepsilon=0.4$,
		whereas the red loop $\Omega_2$ appears at $\varepsilon=0.3$ then disappears at $\varepsilon=0.6$.
		For each $\varepsilon$, the number of connected components and the number of loops are listed underneath.
		\label{fig:pointcloud}}
\end{figure*}

At each timescale $\tau$, we consider mapping $\chi_{\tau}$ from the set $V_{\cG}=\{v_1,v_2,\ldots, v_N\}$ of nodes in $\cG$ to the Euclidean space $\mathbb{R}^N$ such that,
\begin{align}
    \chi_{\tau} : \quad V_{\cG}&\longrightarrow \mathbb{R}^N \nonumber\\
    v_i&\longmapsto \bpg(\tau|i)\quad (i=1,2,\ldots,N).
\end{align}
The mapped point $\bpg(\tau|i)$ of nodes $v_i$ represents the probability on
all nodes at time $\tau$ of a random walker that starts from $v_i$.
Therefore, $\bpg(\tau|i)$ can reflect the interaction between $v_i$
and other nodes at $\tau$-scale, and characterize the structural role of node $v_i$ with multi-resolutions when $\tau$ varies.
The shape of the point cloud $P_{\cG}(\tau)=\{ \bpg(\tau|1), \ldots, \bpg(\tau|N)\}$ provides valuable insights 
into the dyadic and nondyadic interactions between nodes, 
and into the structural property of $\mathcal{G}$ at $\tau$-scale.
Moreover, the distance between two mapped points in $P_{\cG}(\tau)$ is relatively small if there are many paths connecting two original nodes in $\cG$. 
The nodes that belong to the same community or cluster in the network tend to form a group of close points in $P_{\cG}(\tau)$.

Information on the shape of the point cloud can be obtained quantitatively
using the method of persistent homology from computational topology~\cite{edels:2002:persis,zomo:2005:persis,carlsson:2009:topology,edels:2010:topobook}.
The idea is to construct from $P_{\cG}(\tau)$ the $\varepsilon$-scale Vietoris--Rips complex model $\vrip(P_{\cG}(\tau), \varepsilon)$, which is a set of simplices built with a nonnegative threshold $\varepsilon$~\cite{kaczynski:2006:computational}.
Here, every collection of $n+1$ affinely independent points in $P_{\cG}(\tau)$ forms an $n$-simplex in $\vrip(P_{\cG}(\tau), \varepsilon)$ if the pairwise distance between the points is less than or equal to $\varepsilon$.
To build the Vietoris--Rips complex model, we consider a union of balls of radius $\varepsilon/2$ centered at each point in $P_{\cG}(\tau)$~(Fig.~\ref{fig:pointcloud}).
Each simplex is built over a subset of points if the balls intersect between every pair of points.
These simplices can represent the nondyadic interactions of nodes at $\tau$-scale.
In turn, the constructed complex $\vrip(P_{\cG}(\tau), \varepsilon)$ provides information on the topological structure of $P_{\cG}(\tau)$ associated with $\varepsilon$.
Now, starting with $\varepsilon=0$, the complex contains only the $0$-simplices, i.e., the discrete points.
As $\varepsilon$ increases, connections exist between the points, enabling us to obtain
a sequence of embedded complexes called filtration with edges ($1$-simplices), 
and triangular faces ($2$-simplices) are included into the complexes.
Moreover, if $\varepsilon$ becomes considerably large,
all the points gets connected with each other, 
whereby no useful information can be conveyed.

\begin{figure}
		\includegraphics[width=8.2cm]{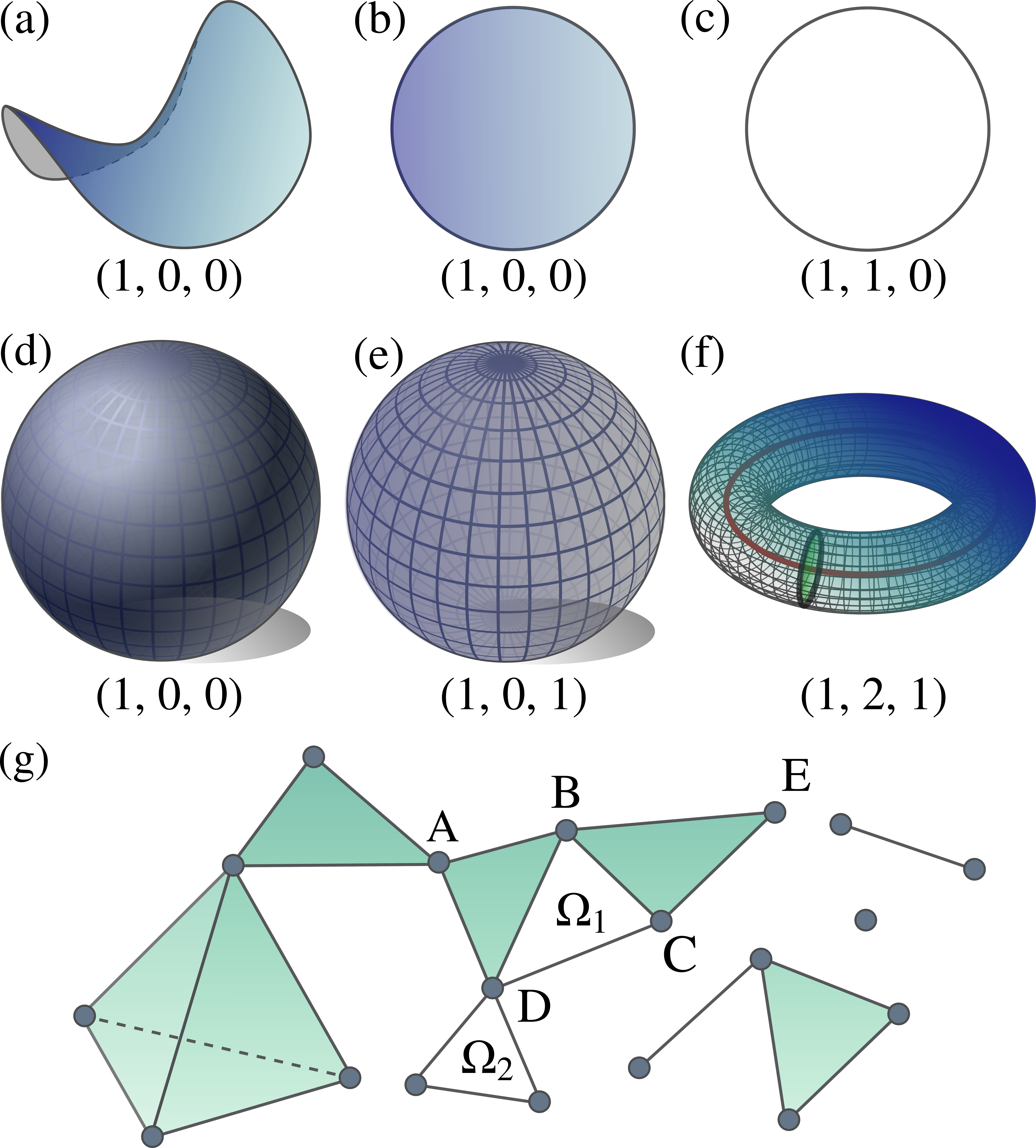}
		\protect\caption{(a)--(f) Sample manifolds with the number of zero-, one-, and two-dimensional holes listed underneath. 
		(a) The connected component is a zero-dimensional hole.
		(b)(c) A one-dimensional hole is obtained by puncturing a disk.
		(d)(e) A two-dimensional hole is obtained by emptying the inside of a ball.
		(f) Two one-dimensional holes are illustrated as two circles in a torus.
		(g) Example of a simplicial complex containing 19 points (0-simplices), 24 edges (1-simplices), eight triangular faces (2-simplices), and one filled tetrahedron (3-simplices). 
		There are two one-dimensional holes $\Omega_1$ and $\Omega_2$ in the complex. In this example, all loops $B\rightarrow C\rightarrow D\rightarrow B$, $A\rightarrow B\rightarrow C\rightarrow D\rightarrow A$, $B\rightarrow E\rightarrow C\rightarrow D\rightarrow B$, and $A\rightarrow B\rightarrow E\rightarrow C\rightarrow D\rightarrow A$ are 1-cycles because they are closed 1-chains, that is, the closed collection of edges (1-simplices).
		Each cycle is not a boundary of any 2-chain (collection of triangular faces);
		thus, it characterizes a one-dimensional hole. Note that these cycles characterize the same hole, $\Omega_1$, because 
		the difference between the two of cycles is the boundary of a 2-chain.
		\label{fig:holes}}
\end{figure}

\begin{figure*}
		\includegraphics[width=15cm]{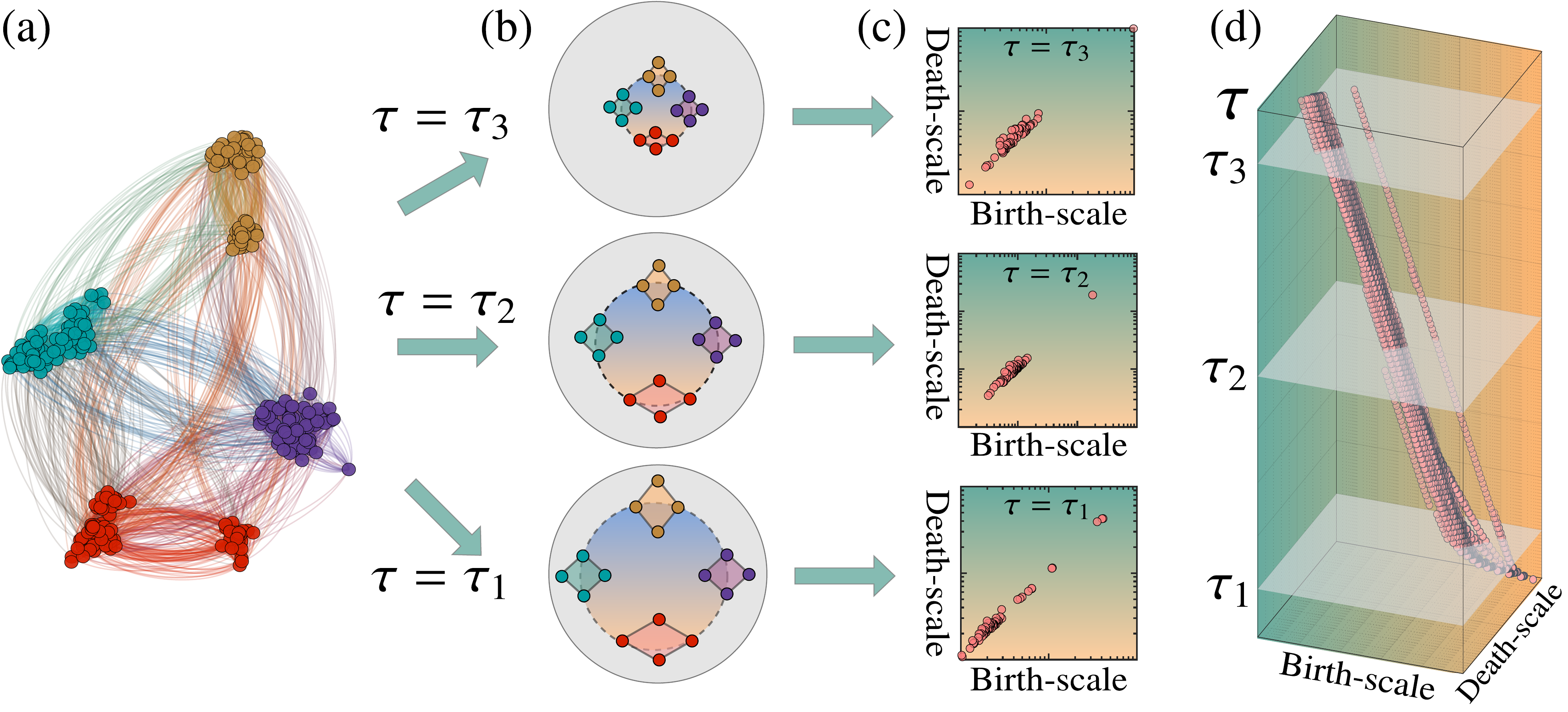}
		\protect\caption{
		(a) An undirected network comprising four clusters with more connections within intra-clusters than between inter-clusters.
		(b) For each $\tau$, the nodes are mapped onto a point cloud such that the distances of the mapped points of the nodes in the same clusters
		are smaller than those between the nodes belonging to different clusters.
		These distances decrease as $\tau$ increases with $\tau_1 < \tau_2 < \tau_3$.
		(c) The topological features at each $\tau$ characterize the shape of the point cloud.
		These features are displayed as a two-dimensional persistence diagram at each $\tau$.
		(d) The scale-variant topological features, i.e., the three-dimensional persistence diagram, 
		are obtained by integrating two-dimensional diagrams at varying $\tau$. 
		The birth-scale and death-scale axes of the diagrams are represented at the logarithmic scale.
		\label{fig:overview}}
\end{figure*}
Persistent homology tracks the variation of topological structures over the filtration.
We refer to the topological structures, i.e., ``holes'' in high-dimensional data, as connected components, tunnels or loops (e.g., a circle of torus), cavities or voids (e.g., the space enclosed by a sphere), and so on.
In persistent homology, a hole is identified via the cycle that surrounds it.
In a given manifold, a cycle is a closed submanifold, 
and a boundary is a cycle that is also the boundary of a submanifold.
Holes correspond to cycles that are not themselves boundaries.
For instance, a disk is a two-dimensional surface with a one-dimensional boundary (i.e., a circle).
If we puncture the disk, we obtain a one-dimensional hole that is enclosed by the circle, which is no longer a boundary [Fig.~\ref{fig:holes}(b)(c)].
Similarly, a filled ball is a three-dimensional object with a two-dimensional boundary (i.e., a surface sphere).
If we empty the inside of the ball, we obtain a two-dimensional hole that is enclosed by the surface sphere, which is no longer a boundary [Fig.~\ref{fig:holes}(d)(e)].
Based on these observations, we can describe and classify holes in the simplicial complex
according to the cycles that enclose holes.
Given a simplicial complex, we define an $n$-chain as a collection of $n$-simplices in the complex.
Therefore, in a simplicial complex, we can define an $n$-cycle as a closed $n$-chain and an $n$-boundary as an $n$-cycle, which is also the boundary of an $(n+1)$-chain.
Here, a $0$-cycle is a connected component, a $1$-cycle is a closed loop, and a $2$-cycle is a shell.
For instance, in Fig.~\ref{fig:holes}(g), all loops $A\rightarrow B\rightarrow D\rightarrow A$, $B\rightarrow C\rightarrow D\rightarrow B$, and $A\rightarrow B\rightarrow C\rightarrow D\rightarrow A$ are 1-cycles because
they are the closed collection of edges (1-simplices).
Furthermore, the loop $A\rightarrow B\rightarrow D\rightarrow A$ is a 1-boundary because it bounds a triangular face (2-simplex).
An $n$-dimensional hole corresponds to an $n$-cycle that is not a boundary of any $(n+1)$-chain in the simplicial complex.
For instance, as illustrated in Fig.~\ref{fig:holes}(g), 
the loops $B\rightarrow C\rightarrow D\rightarrow B$ and $A\rightarrow B\rightarrow C\rightarrow D\rightarrow A$ 
characterize one-dimensional holes because these loops are 1-cycles but are themselves not 1-boundaries.
Moreover, two $n$-cycles characterize the same hole when together they bound an $(n+1)$-chain (i.e., their difference is an $n$-boundary).
Intuitively, the connected components can be considered as zero-dimensional holes, 
the loops and tunnels as one-dimensional holes,
and the cavities and voids as two-dimensional holes.

We consider the emergence and disappearance of holes in the Vietoris--Rips filtration of $P_{\cG}(\tau)$ as topological features for the complex network $\cG$ at $\tau$-scale.
Such features can be observed using multi-set points in a two-dimensional \emph{persistence diagram},
$D^{(2)}_{(l),\tau}(\mathcal{G})$, which is calculated for $l$-dimensional holes.
In this diagram, each point $(b,d)$ denotes a hole 
that appears at the \emph{birth-scale}, $\varepsilon=b$, and disappears at the \emph{death-scale}, $\varepsilon=d$~(see Appendix~\ref{sec:appx:filt}).
Observing the above-defined features, i.e., the two-dimensional persistence diagrams 
with varying $\tau$ can provide insights 
into the variation of topological structures,
thereby reflecting the variation of topological scales in the network.
For instance, the persistence diagrams of zero-dimensional and one-dimensional holes contain information on clusters, connected components, or loops in the point cloud $P_{\cG}(\tau)$, and thus lead to an understanding of the formation of communities and loops in the network at the $\tau$-scale.
We construct scale-variant topological features 
by regarding $\tau$ as a variable parameter rather than as a single fixed value.

In Fig.~\ref{fig:overview}(a), we consider an undirected network that comprises four clusters with more intra-cluster connections than inter-cluster ones.
Pairwise distances of the mapped points of the nodes belonging to the same clusters 
are smaller than the distances between the nodes belonging to different clusters.
These distances decrease as values of $\tau$ increase~[Fig.~\ref{fig:overview}(b)].
In the point cloud, the hole patterns appear with different sizes in different groups of points as $\tau$ varies.
We obtain the scale-variant topological features that reflect the variation of topological scales
by considering the two-dimensional persistence diagrams with the varying $\tau$.
Consider $\tau$ in a set $\mathcal{T}=\{\tau_1,\tau_2,...,\tau_K\}$, 
where $0 < \tau_1<\tau_2<\cdots<\tau_K$ are predefined or sampled values from the continuous domain of timescales.
The scale-variant topological features, i.e., the three-dimensional persistence diagram 
of $l$-dimensional holes for network $\mathcal{G}$, 
are defined by $D^{(3)}_{(l)}(\mathcal{G})=\{(b,d,\tau)\mid (b,d) \in D^{(2)}_{(l),\tau}(\mathcal{G}), \tau \in \mathcal{T}\}$~[Fig.~\ref{fig:overview}(d)].

\subsection{Robustness of scale-variant topological features}

We show that the scale-variant topological features 
are robust with respect to some perturbations of the network.
To describe this robustness, we use the bottleneck distance, $d^{(3)}_{\textup{B}, \xi}$,
a metric structure introduced in Ref.~\cite{tran:2018:variant} 
for comparing three-dimensional persistence diagrams (see Appendix~\ref{sec:appx:stab}).
Herein, $\xi$ is a positive rescaling coefficient introduced to adjust 
the scale difference between the pointwise distance and time.
We consider two undirected networks $\mathcal{G}$ and $\mathcal{H}$ with the same number of nodes.
Based on Refs.~\cite{golub:2012:matrix,chazal:2014:stab}, we can prove that 
the upper limit of the bottleneck distance between 
$D_{(l)}^{(3)}(\cG)$ and $D_{(l)}^{(3)}(\cH)$ is governed
by the matrix 2-norm of the difference between $\bLg$ and $\bLh$ (see Appendix~\ref{sec:appx:stab}):
\begin{align}\label{eqn:bottleneck:stablility}
    d^{(3)}_{\textup{B}, \xi}(D^{(3)}_{(l)}(\cG), D^{(3)}_{(l)}(\cH)) \leq 2\tau_K \Vert \bLg - \bLh \Vert_2.
\end{align}
Herein, $\Vert \boldsymbol{A}\Vert_2$ denotes the matrix 2-norm of matrix $\boldsymbol{A}$.
The inequality of Eq.~\eqref{eqn:bottleneck:stablility} indicates that
our scale-variant topological features are robust with respect to the perturbations 
applied to the random walk Laplacian matrix.
Therefore, these features can be used as discriminative features for characterizing networks.

\subsection{Kernel method for scale-variant topological features}

In the statistical-learning tasks, many learning algorithms require an inner product between the data in the vector form.
Because the space of three-dimensional persistence diagrams is not a vector space,
we deem it not straightforward to use the scale-variant topological features in the statistical-learning tasks.
This problem can be mitigated through the use of a feature map $\Phi$ from the positive-definite kernel,
which maps the scale-variant topological features to a space called \textit{kernel-mapped feature space} $H_b$
where we can define the inner product~\cite{tran:2018:variant}.
In general, choosing the explicit form of mapping a persistence diagram $E$ to $\Phi_E$ in the kernel-mapped feature space is not discernible.
Nonetheless, we can use a kernel function to compute the inner product in the kernel-mapped feature space, leaving the mapping function 
and the kernel-mapped feature space completely implicit.

Given a positive bandwidth $\sigma$ and a positive rescaling coefficient $\xi$ 
introduced to adjust the scale difference between the point-wise distance and time (see Appendix~\ref{sec:appx:param}),
based on Refs.~\cite{reininghaus:mskernel:2015,tran:2018:variant},
we define the kernel $\cK_{\sigma, \xi}$ between two three-dimensional persistence diagrams, $E$ and $F$, as
\begin{align}\label{eqn:kernel}
\cK_{\sigma, \xi}(E, F)=\frac{1}{\sigma\sqrt{2\pi}}\sum_{\bqa\in E\atop\bqb \in F} \left( e^{-\frac{d_{\xi}^2(\bqa, \bqb)}{2\sigma^2}} - e^{-\frac{d_{\xi}^2(\bqa, \bbqb)}{2\sigma^2}} \right),
\end{align}
where $d^2_{\xi}(\bqa, \bqb)=|b_1-b_2|^2+|d_1-d_2|^2+\xi^2|\tau_1-\tau_2|^2$,  $d^2_{\xi}(\bqa, \bbqb)=|b_1-d_2|^2+|d_1-b_2|^2+\xi^2|\tau_1-\tau_2|^2$, with $\bqa=\left(b_1,d_1,\tau_1\right)$ and $\bqb=\left(b_2,d_2,\tau_2\right), \bbqb=\left(d_2,b_2,\tau_2\right)$.
In our experiments, we use the normalized version of the kernel, which is calculated as
\begin{align}\label{eqn:kernel:label}
\cK_{\sigma, \xi}(E, F) \leftarrow \cK_{\sigma, \xi}(E, F)/\sqrt{\cK_{\sigma, \xi}(E, E)\cK_{\sigma, \xi}(F, F)}.
\end{align}
Because Eq.~$\eqref{eqn:kernel}$ and Eq.~$\eqref{eqn:kernel:label}$ define
the positive-definite kernels in the set of three-dimensional persistence diagrams~\cite{tran:2018:variant},
according to Moore--Aronszajn's theorem~\cite{aronszajn:50:reproducing},
there exists a mapping function $\Phi$ such that 
the inner product $\langle \Phi_E, \Phi_F\rangle_{H_b}$ between $\Phi_E$ and $\Phi_F$ in the kernel-mapped feature space $H_b$ is $\cK_{\sigma, \xi}(E, F)$.
Therefore, we can use the explicit form of inner product $\langle \Phi_E, \Phi_F\rangle_{H_b}$ in the statistical-learning tasks.

Furthermore, we can use the above-defined kernel to estimate the transition points with respect to the topological structure in the series of networks $\cG_1, \cG_2, \ldots, \cG_M$.
Consider a collection of diagrams  $\mathcal{D}_{(l)}=\{D_{(l),1}^{(3)},D_{(l),2}^{(3)},\ldots,D_{(l),M}^{(3)}\}$, 
where $D_{(l),i}^{(3)}$ is the three-dimensional persistence diagram of $l$-dimensional holes for network $\cG_i$ ($i=1,2,\ldots,M)$.
Here, we define the transition with respect to the topological structure in $\cG_1, \cG_2, \ldots, \cG_M$ as they abruptly change at given unknown instants (change-points) in $\mathcal{D}_{(l)}$.
We use the kernel change-point detection method~\cite{harchaoui:2009:kernel} to
solve the change-point regression problem with $\Phi_{D^{(3)}_{(l),1}}, \Phi_{D^{(3)}_{(l),2}}, \ldots, \Phi_{D^{(3)}_{(l),M}}$.
Given an index $s$ ($1 < s \leq M)$, we calculate the kernel Fisher discriminant ratio $\kfdr_{M,s}(\mathcal{D}_{(l)})$, 
which is a statistical quantity to measure 
the dissimilarity between two classes assumptively defined by two sets of diagrams 
having index before and from $s$ (see Appendix~\ref{sec:appx:fisher}).
Here, the index $s$ achieving the maximum of $\kfdr_{M,s}(\mathcal{D}_{(l)})$ 
corresponds to the estimated transition point.

\section{Results}

\subsection{Understanding variations of the parameters of network models}

We now investigate how the scale-variant topological features can
reflect variations of the parameters of network models.
We generate networks using Girvan--Newman (GN)~\cite{girvan:2004:benchmark},
Lancichinetti--Fortunato--Radicchi (LFR)~\cite{lanci:2008:lfr,lanci:2009:overlap}, 
Watts--Strogatz (WS)~\cite{watts:1998:collective}, 
Erd\H{o}s--R{\'e}nyi (ER)~\cite{erdos:1959:random}, 
Lancichinetti--Fortunato--Radicchi with hierarchical structure (LFR--H)~\cite{lanci:2009:hierarchical}, and Sales--Pardo (SP)~\cite{sales:2007:hierarchical} models.
We focus on the model parameters that represent the topological scale of these networks,
such as the ratio $r$ between the probability of inter- ($p_{\text{out}}$) 
and intra-community links ($p_{\text{in}}$) (GN),
mixing rate $\mu$ (LFR),
rewiring probability $\beta$ (WS),
pair-link probability $p_{\text{link}}$ (ER),
mixing rate $\mu_{\text{macro}}$ for macrocommunities (LFR--H),
and $\rho$, which estimates the separations between topological scales in the SP model.
The model parameters are varied as 
		$r=p_{\text{out}}/p_{\text{in}}=0.01, 0.02,\ldots, 1.0$;
		$\mu=0.01, 0.02,\ldots, 1.0$;
		$\beta=0.00, 0.01,\ldots, 1.0$; 
		$p_{\text{link}}=0.020, 0.021, \ldots, 0.1$; 
		$\mu_{\text{macro}}=0.01, 0.02, \ldots, 0.2$ and; 
		$\rho=0.05, 0.10, \ldots, 2.0$.
We generate 10 network realizations for each of the models GN, LFR, WS, ER, and SP, 
and 20 network realizations for the LFR--H model at each value of the corresponding model parameter.
There are 128 nodes in the GN, LFR, WS, and ER networks,
300 nodes in each LFR--H network, and 640 nodes in each SP network.

\begin{figure}
		\includegraphics[width=8.5cm]{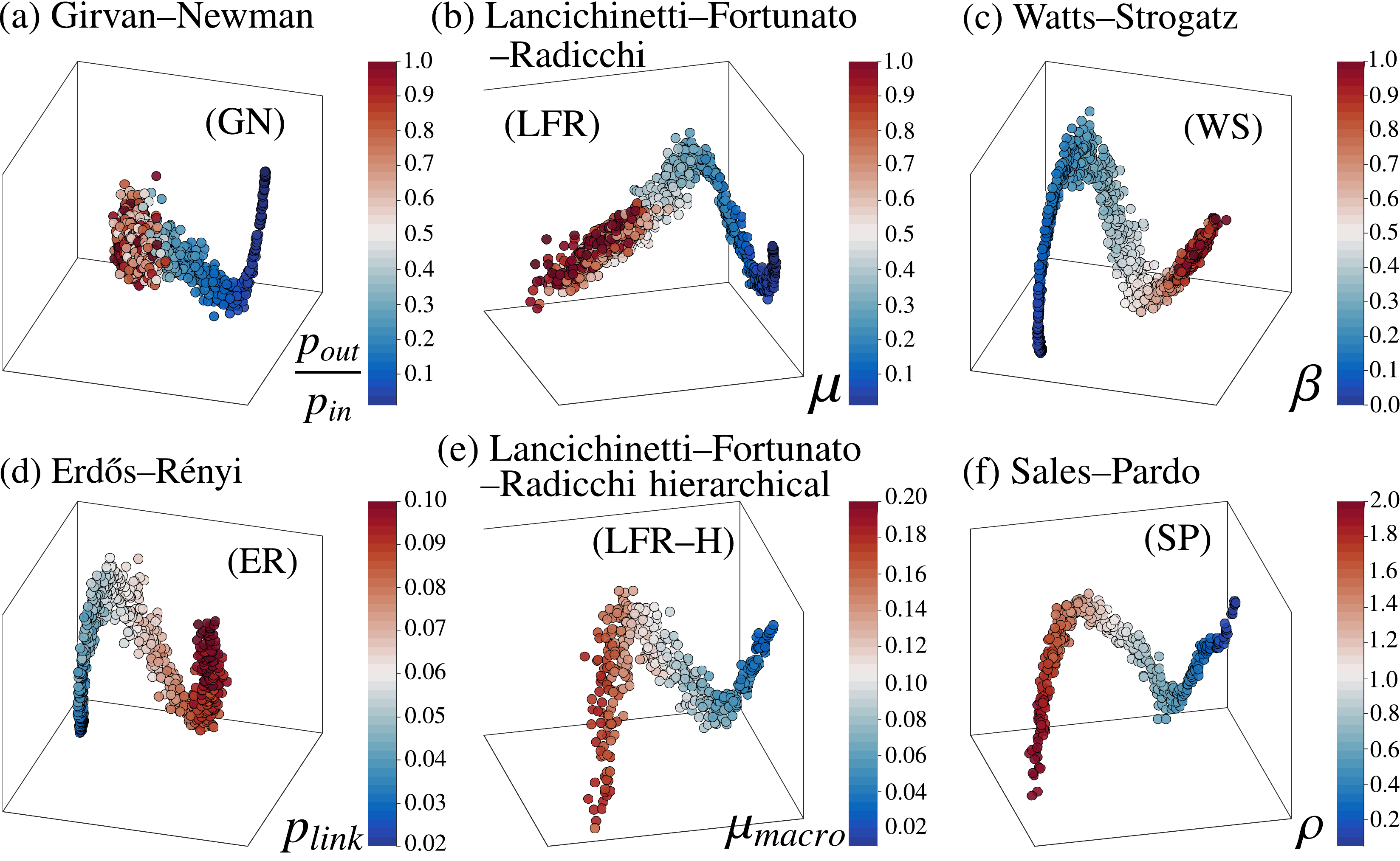}
		\protect\caption{Principal components projection from the kernel-mapped feature space of the scale-variant topological features 
		in each network model.
		Points with different colors represent networks generated from different values of the model parameters.
		Networks are generated from (a) Girvan--Newman (GN), 
		(b) Lancichinetti--Fortunato--Radicchi (LFR), 
		(c) Watts--Strogatz (WS), 
		(d) Erd\H{o}s--R{\'e}nyi (ER), 
		(e) Lancichinetti--Fortunato--Radicchi hiearchical (LFR--H), 
		and (f) Sales--Pardo (SP) models.
		Parameters for these models vary as follows:
		$r=p_{\text{out}}/p_{\text{in}}=0.01, 0.02,\ldots, 1.0$ (GN); 
		$\mu=0.01, 0.02,\ldots, 1.0$ (LFR); 
		$\beta=0.00, 0.01,\ldots, 1.0$ (WS); 
		$p_{\text{link}}=0.020, 0.021, \ldots, 0.1$ (ER); 
		$\mu_{\text{macro}}=0.01, 0.02, \ldots, 0.2$ (LFR--H) and; 
		$\rho=0.05, 0.10, \ldots, 2.0$ (SP).
		\label{fig:kernel-generated}
		}
\end{figure}

\begin{figure}
		\includegraphics[width=8.5cm]{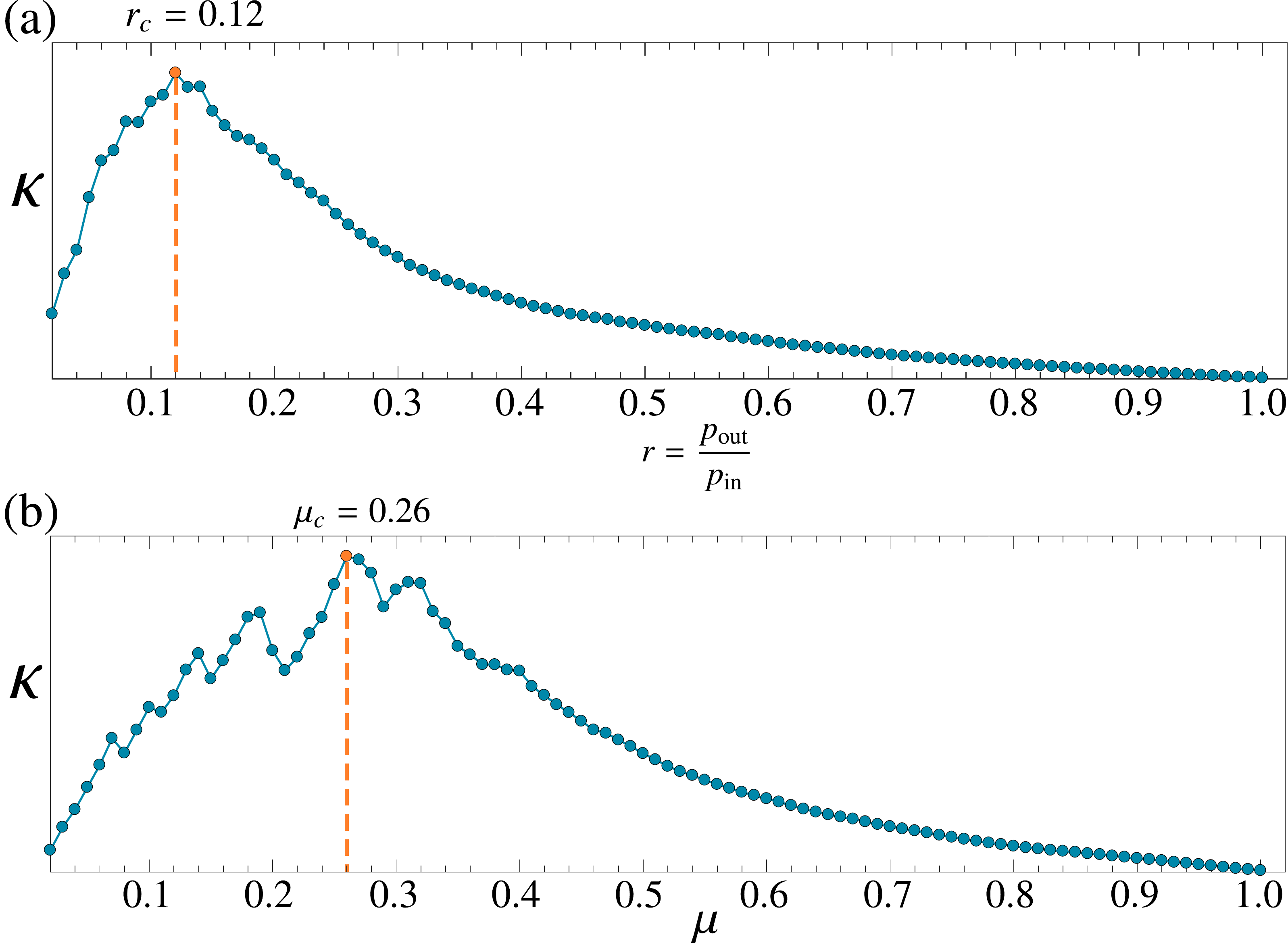}
		\protect\caption{Kernel Fisher discriminant ratio $\kfdr$ estimated for the series of (a) Girvan--Newman networks and (b) Lancichinetti--Fortunato--Radicchi networks.
		The transition point is detected with respect to the topological structure of networks 
		from the series of persistence diagrams for one-dimensional holes obtained when $r=p_{\text{out}}/p_{\text{in}}$ is increased
		as $r_1=0.01, r_2=0.02, \ldots, r_{100}=1.0$ (for Girvan--Newman networks), 
		and $\mu$ is increased as $\mu_1=0.01, \mu_2=0.02, \ldots, \mu_{100}=1.0$ (for Lancichinetti--Fortunato--Radicchi networks).
		The maximum value of $\kfdr$ is marked with the orange point of the dashed line.
		The transition point is the value of the parameter that achieves the maximum value of
		$\kfdr$.
		The transition points are obtained as $r_c=0.12$ (for Girvan--Newman networks) and $\mu_c=0.26$ (for Lancichinetti--Fortunato--Radicchi networks).
		\label{fig:kfdr-GN-LFR}}
\end{figure}

We compute three-dimensional persistence diagrams for one-dimensional holes with 
$\tau_1=1,\tau_2=2,\ldots,\tau_{100}=100$, and
then calculate the kernel defined in Eq.~\eqref{eqn:kernel} for the collection of generated networks in each model.
Figure~\ref{fig:kernel-generated} shows the principal components projections
from the kernel-mapped feature space of each model,
at which the points with different colors represent the networks 
generated from different values of the model parameters.
In WS, ER, LFR--H, and SP models, the scale-variant topological features reflect
a variation of the parameters associated with the topological scales  
mainly that the points located at different positions have different colors [Fig.~\ref{fig:kernel-generated}(c)--(f)].
In GN and LFR models, there are variations in the topological scales of the network 
as $r$ and $\mu$ vary from 0 (four separate groups) to 1 (a purely random graph).
Using the kernel Fisher discriminant ratio calculated for the series of persistence diagrams, we obtain the transition with respect to the topological structure at $r=0.12$ and $\mu=0.26$ for the series of networks obtained at increasing $r$ and $\mu$
~(Fig.~\ref{fig:kfdr-GN-LFR}).
These values correspond to the boundaries between the identifiable phases, 
where parameters can be identified from the kernel-mapped feature space 
and the non-identifiable phases [Fig.~\ref{fig:kernel-generated}(a)--(b)].

\subsection{Identification of network models}
\begin{figure*}
		\includegraphics[width=16cm]{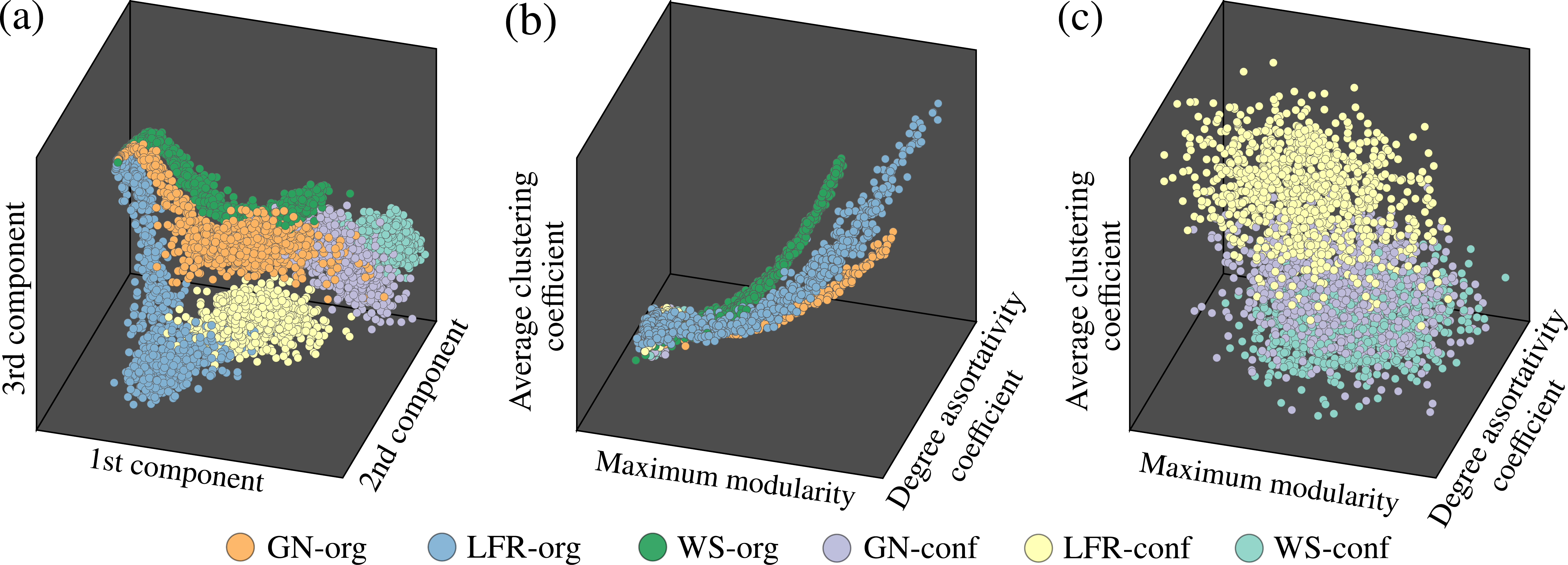}
		\caption{Networks from Girvan--Newman, Lancichinetti--Fortunato--Radicchi, 
		and Watts--Strogatz models are generated with labels denoted by GN-org, LFR-org, and WS-org, respectively; their corresponding configuration networks labels are denoted by GN-conf, LFR-conf, and WS-conf.
		(a) The kernel principal components projection of the scale-variant topological features for these networks. 
		(b)(c) Variation of high-order features, i.e., degree assortativity coefficient, maximum modularity, and average clustering coefficient for (b) all generated networks, and (c) configuration networks.
		The different colors represent the networks generated from different models.
		\label{fig:kernel-PCA}}
\end{figure*}
Here we show that the scale-variant topological features can classify the networks generated from different models, 
even if they have similar global statistical measures.
We study the configuration model in Ref.~\cite{newman:2001:random}, 
which generates random networks (known as \textit{configuration networks}) having the same sequences of node degrees as a given network.
The labels of the networks generated from GN, LFR, and WS models 
are denoted by GN-org, LFR-org, and WS-org, respectively,
while their corresponding configuration networks labels are denoted by 
GN-conf, LFR-conf, and WS-conf.
We compute the three-dimensional persistence diagrams for one-dimensional holes of these networks
with timescale values $\tau_1=1,\tau_2=2,\ldots,\tau_{100}=100$.
Accordingly, we calculate the kernel for these diagrams, then perform
three-dimensional projections of the principal components 
from the kernel-mapped feature space [Fig.~\ref{fig:kernel-PCA}(a)].
Here, points with different colors represent networks generated from different models.
In Fig.~\ref{fig:kernel-PCA}(a), the points appear to be distinguishable by their colors,
thus, we can conclude that the scale-variant topological features can characterize the differences
with respect to the topological structure between networks, 
and even between configuration networks generated from different models.

While the node degree distribution in a configuration network is the same as the given network,
the topological correlations between the nodes are destroyed.
Therefore, we investigate conventional
higher-order features of the network, such as 
the degree assortativity coefficient, the average clustering coefficient,
and the maximum modularity obtained via Louvain heuristic~\cite{blondel:2008:unfold,girvan:2004:benchmark}.
Figure~\ref{fig:kernel-PCA}(b) highlights the variation of these features in our generated networks.
Specifically in Fig.~\ref{fig:kernel-PCA}(b), the points with corresponding labels GN-org, LFR-org, and WS-org appear to
be distinguishable with others, thus,
it becomes easy to distinguish between networks generated from different models 
and between a given network with its corresponding configuration network.
However, if we look at the variation of these features for configuration networks
[Fig.~\ref{fig:kernel-PCA}(c)], we note that
the conventional higher-order features of the network cannot capture the apparent differences 
in topological structure between the configuration networks, 
even when their corresponding original networks are generated from different mechanics models.
In contrast with this observation and as highlighted in Fig.~\ref{fig:kernel-PCA}(a), 
the scale-variant topological features can provide a better 
representation of the topological structure of networks.

\begin{figure}
		\includegraphics[width=9.0cm]{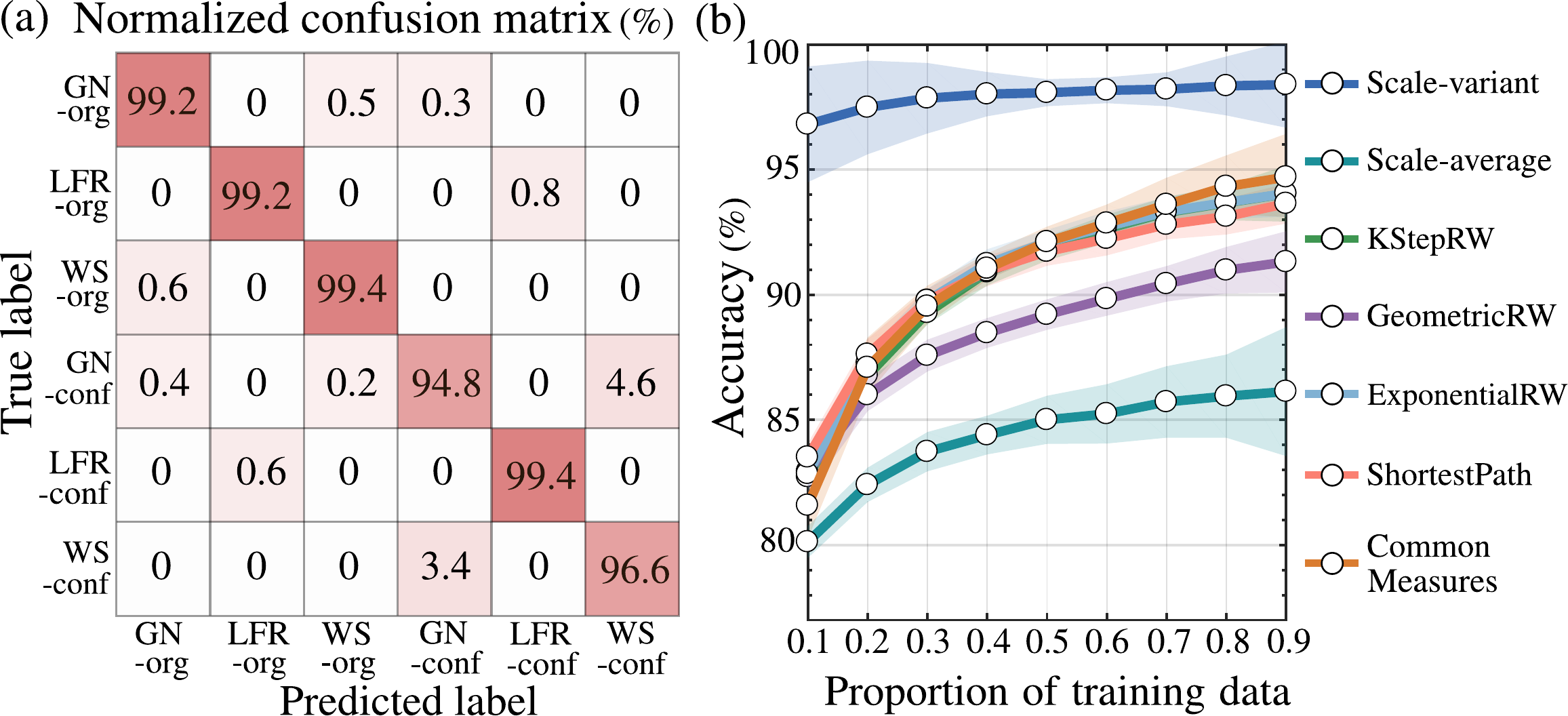}
		\protect\caption{Classification of networks generated from Girvan--Newman, Lancichinetti--Fortunato--Radicchi, 
		and Watts--Strogatz models, with labels denoted by GN-org, LFR-org, and WS-org, respectively; their corresponding configuration networks labels are denoted by GN-conf, LFR-conf, and WS-conf.
		(a) Average normalized confusion matrix of the scale-variant method over 100 random train-test splits of the data.
		The 10 networks generated at each value of the model parameters are split into two, with
		five networks for training and the other five for testing.
		(b) Average accuracies (\%) of the classification methods 
		over 100 random train-test splits at each proportion of the training data (bold lines).
		The shaded areas indicate the confidence intervals of 
		one standard deviation calculated using the same ensemble of runs.
		\label{fig:kernel-joint}}
\end{figure}

Accordingly, we quantify to what extent the scale-variant topological features identify the networks generated from different models.
We employ the scale-variant method, which uses the scale-variant topological features to classify networks into six labels, namely, 
GN-org, LFR-org, WS-org, GN-conf, LFR-conf, and WS-conf.
We randomly split 10 networks generated at each value of the model parameters into two, i.e., 
five networks for training and five for testing, and apply
the support vector machine~\cite{bishop:2006:prm} for classification in the kernel-mapped feature space. 
Figure~\ref{fig:kernel-joint}(a) depicts the average normalized confusion matrix 
over 100 random splits, where the row and column labels are the predicted and true labels, respectively.
Figure~\ref{fig:kernel-joint}(a) shows a reasonably high accuracy for identifying the networks generated from different models with the following labels: GN-org ($99.2\%$), LFR-org ($99.2\%$), WS-org ($99.4\%$), GN-conf ($94.8\%$), LFR-conf ($99.4\%$), and WS-conf ($96.6\%$).
This result demonstrates that the scale-variant topological features 
can reflect well on the behaviors of these network models.

To highlight the benefits of the scale-variant method, we compare it with the other conventional methods using common network measures~\cite{freeman:1978:centrality,latora:2001:efficient,newman:2002:assortative,costa:2007:characterization,brandes:2008:variants}, 
well-recognized graph kernels~\cite{sugiyama:2017:graphkernels},
and topological features calculated at an average fixed topological scale.
We describe the common network measures in Appendix~\ref{sec:appx:common} as well as 
the graph kernels that are based on random walks (KStepRW, GeometricRW, ExponentialRW)~\cite{kashima:2003:marginalized,gartner:2003:graph}, 
paths (ShortestPath)~\cite{borgwardt:2005:shortest}, 
limited-sized subgraphs (Graphlet)~\cite{borgwardt:2007:efficient},
and subtree patterns (Weisfeiler--Lehman~\cite{shervashidze:2011:weisfeiler}) in Appendix~\ref{sec:appx:graphker}.
Moreover, we consider two variations of topological features evaluated at an average fixed topological scale
to show the advantages of using variable timescales.
Also, instead of using a particular timescale, we use the scale-average and the scale-norm-average methods to preserve the geometrical persistence of the point cloud.
The former uses the topological features extracted from the average distance matrix $\bDelta_{\text{avg}}=(1/K)\sum_{i=1}^K\bDelta_{\tau_i}$,
whereas the latter uses the features from the average normalized distance matrix $\Tilde{\bDelta}_{\text{avg}}=(1/K)\sum_{i=1}^K\Tilde{\bDelta}_{\tau_i}$~\cite{domenico:2017:diffusion}. 
Herein, $\bDelta_{\tau_i}$ denotes the distance matrix of pairwise Euclidean distances between points in $P_{\cG}(\tau_i)$,
whereas $\Tilde{\bDelta}_{\tau_i}$ is obtained by dividing $\bDelta_{\tau_i}$ by its maximum element.
We randomly split the 10 networks generated at each value of the model parameters
into proportions for training and for testing,
and employ the support vector machine as the classifier 
to both the common network measures and the kernel-mapped feature space.
We compute the average classification accuracy over 100 random splits 
at different proportions of the training data.
Figure~\ref{fig:kernel-joint}(b) depicts
the performance of the methods with accuracies greater than 70\%, 
mainly illustrating that the scale-variant method outperforms the other methods 
in terms of classification accuracy.
Moreover, the scale-variant method is shown to achieve approximately 97\% of accuracy,
even with a small size of the training dataset, e.g., only 10$\%$ of all the data, 
whereas the other methods yielded accuracies of at most 84\%.
These results validate the effectiveness and the reliability of our scale-variant method
in capturing the differences between network structures.
The source code used in our experiments is publicly available on GitHub~\cite{gitsource:scale}.

\subsection{Classification of the real-world network data}
Next, we apply the scale-variant topological features to the classification of
chemoinformatics network datasets (MUTAG, BZR, COX2, DHFR, FRANKENSTEIN, NCI1, NCI109),
bioinformatics dataset (PROTEIN),
and large real-world social network datasets, such as
movie collaboration networks (IMDB--BINARY, IMDB--MULTI),
scientific collaboration networks (COLLAB),
and networks obtained from online discussion threads on Reddit (REDDIT--BINARY, REDDIT--MULTI--5K)~\cite{debnath:1991:structure,sutherland:2003:spline,kazius:2005:derivation,borgwardt:2005:protein,wale:2008:comparison,orsini:2015:graph,yanardag:2015:deep,kristian:2016:benchmark}.
The aggregate statistics for these datasets is provided in Table~\ref{tab:stats}.
We compute three-dimensional persistence diagrams with $\tau_1=1,\ldots,\tau_{50}=50$, 
and use the multiple kernel learning method~\cite{cortes:2012:center} to learn the linear combination of the normalized kernels for zero-dimensional and one-dimensional holes.
Subsequently, we compare our scale-variant method with methods employing the common network measures and the scale-average and scale-norm-average methods.
Likewise, we compare the scale-variant method with many state-of-the-art algorithms in classifying graphs and networks as follows:
(i) random walk kernels based on matching walks in two graphs (KStepRW, GeometricRW, ExponentialRW)~\cite{kashima:2003:marginalized,gartner:2003:graph}, 
(ii) the shortest path kernel (ShortestPath)~\cite{borgwardt:2005:shortest}, 
(iii) the graphlet count kernel (Graphlet)~\cite{borgwardt:2007:efficient}, 
(iv) the Weisfeiler--Lehman subtree kernel (Weisfeiler--Lehman)~\cite{shervashidze:2011:weisfeiler},
(v) the deep graph kernel (DGK)~\cite{yanardag:2015:deep}, 
(vi) the PATCHY-SAN convolutional neural network (PSCN)~\cite{niepert:2016:learning}, 
and (vii) the graph kernel based on return probabilities of random walks (RetGK)~\cite{zhang:2018:retgk}.
Here, in order to make a fair comparison with these methods, as presented in the literature Ref.~\cite{zhang:2018:retgk}, we apply the support vector machine~\cite{bishop:2006:prm} as the classifier in the kernel-mapped feature space.
Moreover, we perform 10-fold cross-validations, where a single 10-fold is created by randomly shuffling the dataset, and then splitting it into 10 different parts (folds) of equal size. 
In every single 10-fold, we use nine folds for training and one for testing and averaging of the classification accuracy of the test set obtained throughout the folds. 
To reduce the variance of the accuracy due to the splitting of data, 
we repeat the whole process of cross-validation for 10 times,
and then report the average and standard deviation of the classification accuracies.

\begin{table*}
	\caption{\label{tab:stats} Summary statistics of the real-world network datasets.}
	\begin{ruledtabular}
    \begin{tabular}{|l|c|c|c|c|c|c|} 
	\st{Dataset\\~} & \st{Type of\\networks} &\st{Number of\\networks} &\st{Number of\\classes} &\st{Number of networks\\in each class} &\st{Avg. number\\of nodes} &\st{Avg. number\\of edges}\\ \hline
	MUTAG        & Chemoinformatics & 188  & 2 & (63,125)         &  17.93 & 19.79 \\
	BZR          & Chemoinformatics & 405  & 2 & (319,  86)       &  35.75 & 38.36 \\
	COX2         & Chemoinformatics & 467  & 2 & (365,  102)      &  41.22 & 43.45 \\
	DHFR         & Chemoinformatics & 756  & 2 & (295,  461)      &  42.43 & 44.54 \\
	FRANKENSTEIN & Chemoinformatics & 4337 & 2 & (2401, 1936)     &  16.90 & 17.88 \\
	NCI1         & Chemoinformatics & 4110 & 2 & (2053, 2057)     &  29.87 & 32.30 \\
	NCI109       & Chemoinformatics & 4127 & 2 & (2048, 2079)     &  29.68 & 32.13 \\
	PROTEINS     & Bioinformatics & 1113 & 2 & (663,  450)      &  39.06 & 72.82 \\
	IMDB--BINARY  & Social & 1000 & 2 & (500,  500)      &  19.77 & 96.53 \\
	IMDB--MULTI   & Social & 1500 & 3 & (500,  500, 500) &  13.00 & 65.94 \\
	COLLAB   & Social & 5000 & 3 & (2600, 775, 1625) &  74.49 & 2457.78 \\
	REDDIT--BINARY   & Social & 2000 & 2 & (1000, 1000) &  429.63 & 497.75 \\
	REDDIT--MULTI--5K   & Social & 4999 & 5 & (1000, 1000, 1000, 1000, 999) &  508.52 & 594.87 \\
    \end{tabular}
	\end{ruledtabular}
\end{table*}

\begin{table*}
	\caption{\label{tab:social:acc} Average and standard deviation (mean$\pm$sd) of the classification accuracy (\%) for social network datasets IMDB--BINARY, IMDB--MULTI, COLLAB, REDDIT--BINARY, and REDDIT--MULTI--5K. 
	These social network datasets contain networks that do not have information such as labels and attributes of nodes. In each dataset, the best and the second-best scores are colored in dark pink and light pink, respectively. The notation $(*)$ indicates that the kernel computation with the implementation in~\cite{sugiyama:2017:graphkernels} is not completed after 72h.}
		\begin{ruledtabular}
	    \begin{tabular}{lccccc} 
		\st{Method\\~} & \st{IMDB--BINARY\\~} &\st{IMDB--MULTI\\~} &\st{COLLAB\\~} &\st{REDDIT--\\BINARY~} &\st{REDDIT--\\MULTI--5K~} \\ \hline
		Scale-variant  & \rst{74.2 $\pm$ 0.9} & \rst{49.9 $\pm$ 0.3} & \rnd{79.6 $\pm$ 0.3} & \rnd{87.8 $\pm$ 0.3} & 53.1 $\pm$ 0.2\\
		Scale-average  & 67.7 $\pm$ 0.8 & 44.9 $\pm$ 0.4 & 71.4 $\pm$ 0.1 & 79.8 $\pm$ 0.3 & 51.5 $\pm$ 0.2\\
		Scale-norm-average  & 70.2 $\pm$ 0.7 & 44.9 $\pm$ 0.4 & 62.6 $\pm$ 0.1 & 73.9 $\pm$ 0.2 & 48.7 $\pm$ 0.3\\
		CommonMeasures  & \rnd{72.0 $\pm$ 0.2} & 44.9 $\pm$ 0.3 & 75.2 $\pm$ 0.1 &  85.7 $\pm$ 0.3  &  \rst{56.6 $\pm$ 0.2} \\
		KStepRW  & 60.0 $\pm$ 0.8 & 43.8 $\pm$ 0.7 &  $(*)$  &  $(*)$  &  $(*)$ \\
		GeometricRW  & 67.0 $\pm$ 0.8 & 45.2 $\pm$ 0.4 &  $(*)$  &  $(*)$  &  $(*)$ \\
		ExponentialRW  & 65.2 $\pm$ 1.1 & 43.1 $\pm$ 0.4 &  $(*)$  &  $(*)$  &  $(*)$ \\
		ShortestPath  & 58.2 $\pm$ 1.0 & 42.0 $\pm$ 0.6 & 58.5 $\pm$ 0.2 &  81.9 $\pm$ 0.1  & 49.0 $\pm$ 0.1 \\
		Graphlet  & 65.9 $\pm$ 1.0 & 43.9 $\pm$ 0.4 & 72.8 $\pm$ 0.3 & 77.3 $\pm$ 0.2 & 41.0 $\pm$ 0.2\\
		Weisfeiler--Lehman  & 70.8 $\pm$ 0.5 & \rnd{49.8 $\pm$ 0.5} & 74.8 $\pm$ 0.2 & 68.2 $\pm$ 0.2 & 51.2 $\pm$ 0.3\\
		DGK  & 67.0 $\pm$ 0.6 & 44.6 $\pm$ 0.5 & 73.1 $\pm$ 0.3 & 78.0 $\pm$ 0.4 & 41.3 $\pm$ 0.2\\
		PSCN  & 71.0 $\pm$ 2.3 & 45.2 $\pm$ 2.8 & 72.6 $\pm$ 2.2 & 86.3 $\pm$ 1.6 & 49.1 $\pm$ 0.7\\
		RetGK  & 71.9 $\pm$ 1.0 & 47.7 $\pm$ 0.3 & \rst{81.0 $\pm$ 0.3} & \rst{92.6 $\pm$ 0.3} & \rnd{56.1 $\pm$ 0.5} \\
        \end{tabular}
        \end{ruledtabular}
\end{table*}

\begin{table*}
	\caption{\label{tab:bio:acc} Average and standard deviation (mean$\pm$sd) of the classification accuracy (\%) for chemoinformatics and bioinformatics datasets MUTAG, BZR, COX2, DHFR, FRANKENSTEIN, PROTEINS, NCI1, and NCI109.
	Presented is only a comparison of the methods using the connectivity  between nodes.
		In each dataset, the best and the second-best scores are colored in dark pink and light pink, respectively.}
		\begin{ruledtabular}
	    \begin{tabular}{lcccccccc} 
		\st{Method\\~} & \st{MUTAG\\~} &\st{BZR\\~} &\st{COX2\\~} &\st{DHFR\\~} &\st{FRANKEN\\STEIN}  &\st{PROTEINS\\~} &\st{NCI1\\~}       &\st{NCI109\\~} \\ \hline
		Scale-variant  & \rst{88.2$\pm$1.0} & \rnd{85.9$\pm$0.9} & \rst{78.4$\pm$0.4} & \rst{78.8$\pm$0.7} & \rst{69.0$\pm$0.2} & \rnd{72.6$\pm$0.4} & \rst{71.3$\pm$0.4} & \rst{69.8$\pm$0.2}  \\
      
        Scale-average & 83.0$\pm$1.3 & 78.9$\pm$0.4 & 78.2$\pm$0.0 & 66.9$\pm$0.5 &	61.3$\pm$0.2 & 70.8$\pm$0.2 & 66.5$\pm$0.2 & \rnd{65.8$\pm$0.2}  \\
        
       Scale-norm-average & 84.6$\pm$0.9 & 81.7$\pm$0.2 & 78.2$\pm$0.0 & 61.0$\pm$0.0 &	60.2$\pm$0.1 & 71.7$\pm$0.4 & 65.2$\pm$0.1 & 65.7$\pm$0.1  \\ 
      CommonMeasures & \rnd{84.9$\pm$0.3} & 82.8$\pm$0.3 & 78.2$\pm$0.0 & 71.1$\pm$0.6 & 62.0$\pm$0.2 & \rst{75.3$\pm$0.3} & \rnd{67.8$\pm$0.3} & 65.4$\pm$0.1  \\

        KStepRW   & 81.8$\pm$1.3 & \rst{86.5$\pm$0.5} & 78.0$\pm$0.1 & 73.3$\pm$0.4 &	\rnd{65.4$\pm$0.2} & 71.8$\pm$0.1 & 51.7$\pm$0.7 & 50.4$\pm$0.0  \\
        
      GeometricRW   & 82.9$\pm$0.5 & 79.2$\pm$0.4 & 78.2$\pm$0.0 & 71.4$\pm$1.9 &	55.4$\pm$0.1 & 72.2$\pm$0.1 & 62.6$\pm$0.0 & 63.2$\pm$0.0\\
      
       ExponentialRW  & 83.0$\pm$0.5 & 79.5$\pm$0.5 & 78.2$\pm$0.0 & 74.6$\pm$0.3 &	55.4$\pm$0.1 & 72.2$\pm$0.1 & 62.7$\pm$0.1 & 63.2$\pm$0.1  \\
       
     ShortestPath   & 81.8$\pm$0.9 & 85.6$\pm$0.6 & 78.1$\pm$0.1 & 73.2$\pm$0.5 &	63.8$\pm$0.1 & 72.0$\pm$0.3 & 64.2$\pm$0.1 & 61.1$\pm$2.0 \\

     Graphlet  & 83.0$\pm$0.3 & 78.8$\pm$0.0 & 78.2$\pm$0.0 & 61.0$\pm$0.0 &	55.4$\pm$0.0 & 70.6$\pm$0.1 & 62.4$\pm$0.2 & 62.1$\pm$0.1  \\
        
     Weisfeiler--Lehman   & 83.8$\pm$0.8 & 84.0$\pm$1.2 & \rnd{78.3$\pm$0.2} & \rnd{77.2$\pm$0.6} &	62.3$\pm$1.2 & 71.3$\pm$0.5 & 63.2$\pm$0.1 & 63.6$\pm$0.1 \\
        \end{tabular}
        \end{ruledtabular}
\end{table*}
The social network datasets contain networks that do not have information, such as labels and attributes of nodes.
For movie collaboration datasets (IMDB--BINARY, IMDB--MULTI), collaboration ego-networks are generated for each actor (actress).
In each network, two nodes representing the actors or actresses are connected when they appear in the same movie.
The task is to identify whether a given ego-network of an actor (actress) belongs to one of the predefined movie genres.
For scientific collaboration dataset (COLLAB), collaboration ego-networks are generated for different researchers, 
with the objective of determining whether the collaboration network of a researcher 
belongs to one of the research fields
as High Energy Physics, Condensed Matter Physics, or Astro Physics.
For Reddit datasets, each network is generated from an online discussion thread where nodes correspond to users, and edges correspond to the responses between users.
Here, the task is to identify whether a given network belongs to a
\textit{question/answer}-based community or a \textit{discussion}-based community (REDDIT--BINARY), 
or one of five predefined subreddits (REDDIT--MULTI--5K).
Table~\ref{tab:social:acc} presents the average accuracies along with their standard deviations over ten 10-folds.
The results for Weisfeiler--Lehman kernel, DGK kernel, PSCN, and RetGK kernel are taken from Ref.~\cite{zhang:2018:retgk}.
Specifically in Table~\ref{tab:social:acc}, the best and the second-best average accuracy scores for each social network dataset are colored dark pink and light pink, respectively.
For the social network datasets, 
the scale-variant method either is comparable or outperforms the state-of-the-art classification methods.

For the chemoinformatics network datasets, we predict the function classes of chemical compounds in chemoinformatics.
Here, molecules are represented as small networks with nodes as atoms and edges as covalent bonds.
For the bioinformatics dataset (PROTEINS), proteins are represented as networks, where the nodes are secondary structure elements 
and the edges represent the neighborhood within the 3-D structure or along the amino acid chain. 
We aim to classify the function class membership of the protein sequences into enzymes and non-enzymes.
Note that these chemoinformatics and bioinformatics network datasets 
contain information on the labels and attributes of the nodes,
which is leveraged in DGK, PSCN, and RetGK methods.
For a fair comparison of characterizing the structure of networks,
we present in Table~\ref{tab:bio:acc} the average accuracies and standard deviations 
of the methods that only use the connectivity between nodes.
In the table, the best and the second-best average accuracy scores for each dataset are colored dark pink and light pink, respectively.
Here, on average, the scale-variant method outperforms all the other methods,
and offers the best results for six of the eight datasets and the second-best result for two more.
Further, the classification accuracies of the scale-variant method on MUTAG, FRANKENSTEIN, NICI1, and NCI109 datasets are at least two percentage points higher than those of the best baseline algorithms.
These results suggest that
the scale-variant method can be considered as an effective approach 
in classifying real-world network data.

\subsection{Detection of transition points in the time-evolving gene regulatory network}
\begin{figure}
		\includegraphics[width=8.5cm]{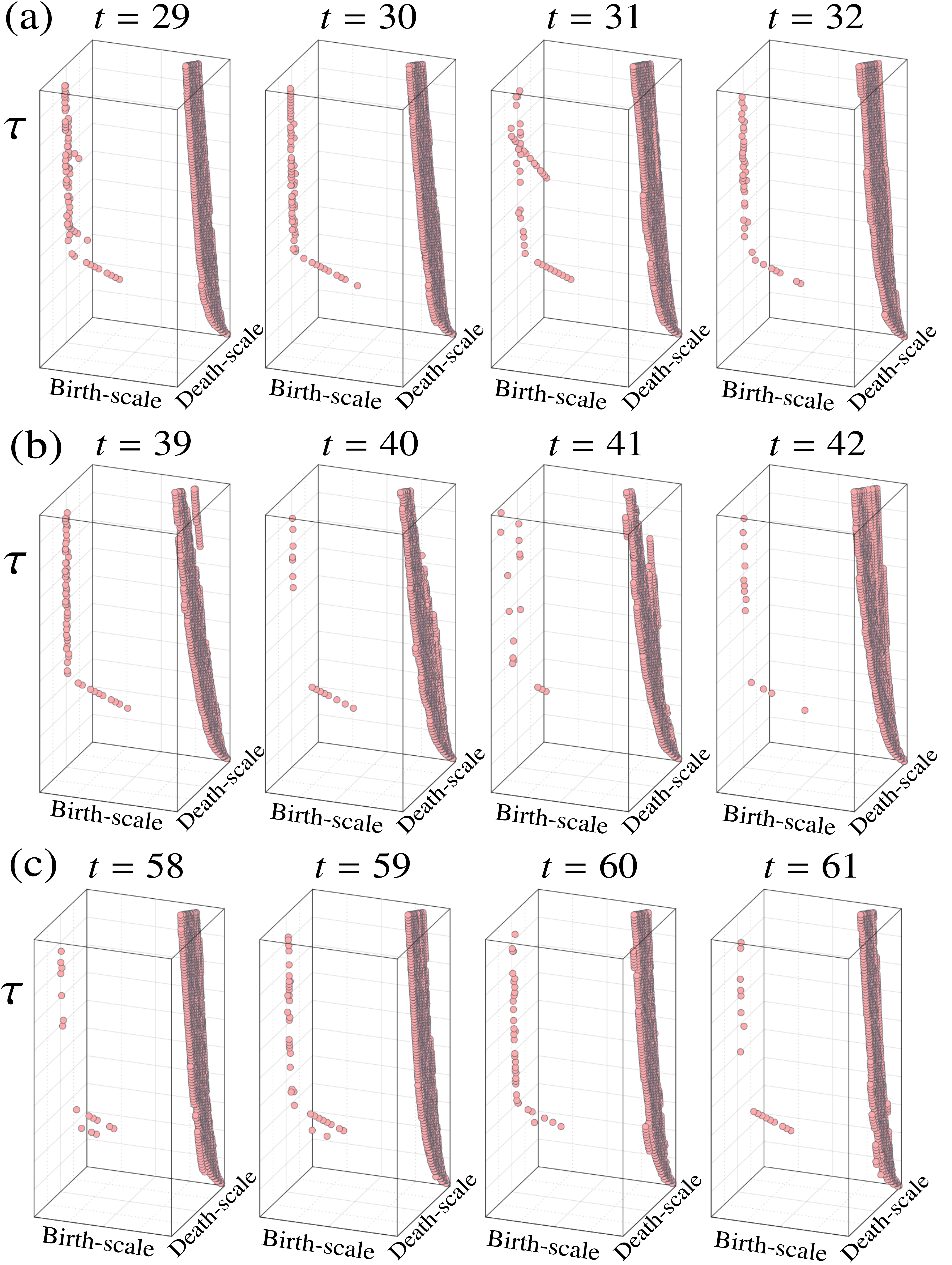}
		\protect\caption{Three-dimensional persistence diagrams of one-dimensional holes for the \textit{Drosophila melanogaster} gene regulatory networks spanning from (a) $t=29$ to $t=32$, (b) $t=39$ to $t=42$, and (c) $t=58$ to $t=61$.
		The birth-scale and the death-scale axes of the diagrams are represented at the logarithmic scale.
		\label{fig:dros:diagram}}
\end{figure}

\begin{figure}
		\includegraphics[width=9cm]{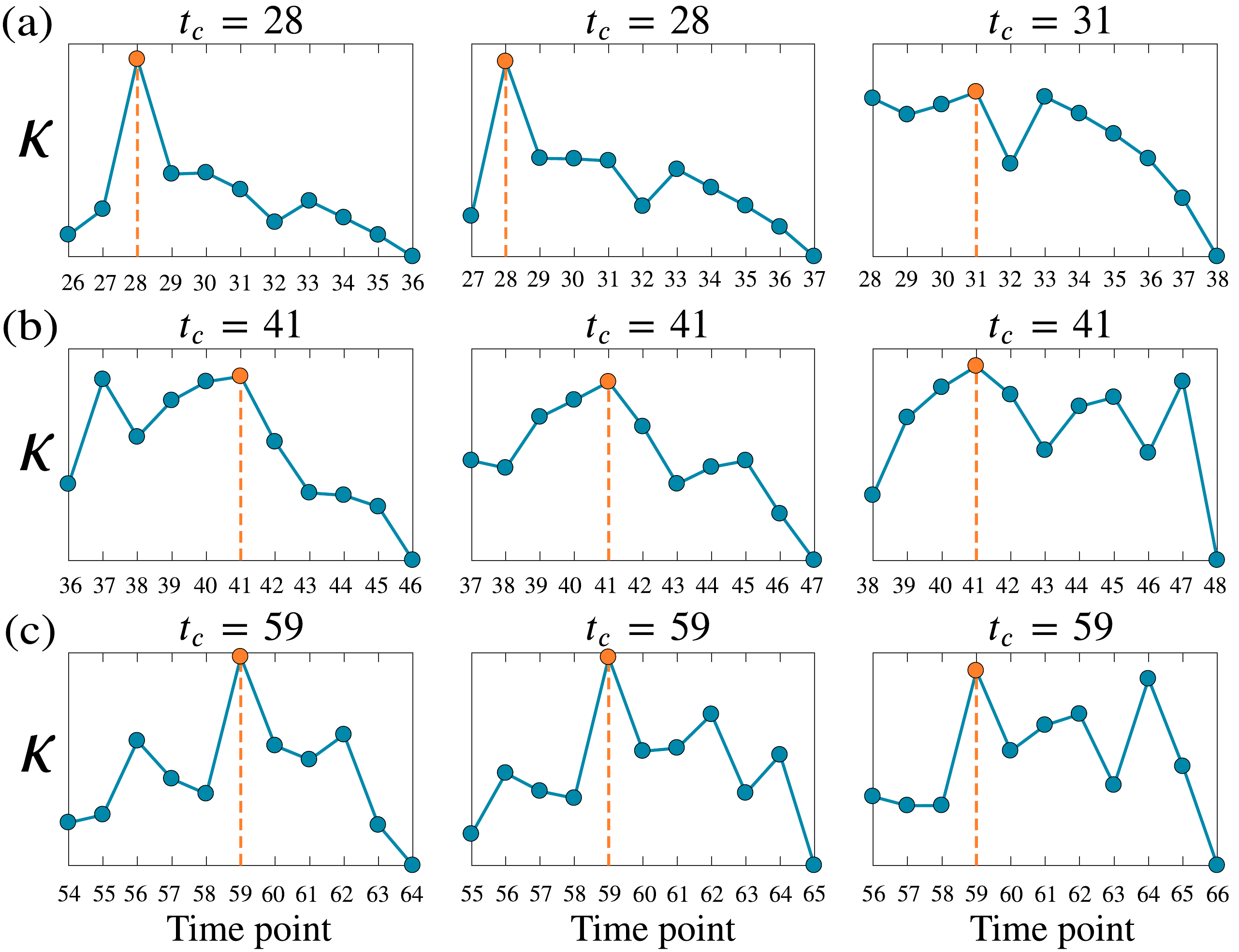}
		\protect\caption{Kernel Fisher discriminant ratio $\kfdr$ calculated from the three-dimensional persistence diagrams
		of one-dimensional holes for the time-evolving \textit{Drosophila melanogaster} gene regulatory networks.
		Transition time points are detected
		in the sliding windows spanning between two different developmental stages.
		In each window, the transition time point $t_c$ is the time index of the maximum $\kfdr$ value marked with the orange point of the dashed line.
		(a) Windows from the embryonic stage to the larval stage with time points $26\rightarrow 36$ ($t_c=28$), $27\rightarrow 37$ ($t_c=28$), and $28\rightarrow 38$ ($t_c=31$).
		(b) Windows from the larval stage to the pupal stage with time points $36\rightarrow 46$ ($t_c=41$), $37\rightarrow 47$ ($t_c=41$) and $38\rightarrow 48$ ($t_c=41$).
		(c) Windows from the pupal stage to the adulthood stage with time points $54\rightarrow 64$ ($t_c=59$), $55\rightarrow 65$ ($t_c=59$) and $56\rightarrow 66$ ($t_c=59$).
		\label{fig:kfdr}}
\end{figure}

We apply the scale-variant topological features to detect transition points 
between the developmental stages of \textit{Drosophila melanogaster} in the
time-evolving gene regulatory networks.
Particularly, we use a genome-wide microarray profiling, which shows the expression patterns 
of 4028 genes simultaneously measured during the developmental stages of \textit{Drosophila melanogaster}~\cite{arbeitman:2002:gene}.
Herein, 66 time points are chosen from the full developmental cycle: 
embryonic stage (1--30), larval stage (31--40), pupal stage (41--58), and adulthood stage (59--66)~\cite{zhang:2017:finding}.
We use kernel reweighted logistic regression method~\cite{song:2009:keller} to reconstruct the time-evolving networks for 588 genes, 
which are known to be related to the developmental process based on their gene ontologies.
Therefore, the networks are reconstructed via logistic regression using only binary information, i.e., activation or non-activation of gene expression data.
For the likelihood being maximized for network inference, a kernel weight function is employed to obtain the dynamic networks structures with smooth transition at adjacent time points~\cite{song:2009:keller}.

We use the scale-variant topological features to detect the transition points with respect to the topological structure of the constructed time-evolving networks.
Consequently, the detected transition points agree with the transition points in the dynamics of \textit{Drosophila melanogaster}.
Here, the  change points between the developmental stages of \textit{Drosophila melanogaster} chosen in the experiment
are referred to as the transition points in the dynamics: $t_1=31$ between the embryonic stage and the larval stage, 
$t_2=41$ between the larval stage and the pupal stage, 
and $t_3=59$ between the pupal stage and the adulthood stage.
For the network of each time point, three-dimensional persistence diagrams are computed for one-dimensional holes with $\tau_1=1,\ldots,\tau_{100}=100$.
Figure~\ref{fig:dros:diagram} shows the examples of the diagrams for networks spanning from (a) $t=29$ to $t=32$, (b) $t=39$ to $t=42$, and (c) $t=58$ to $t=61$.
Here, note the transformation of the patterns in scale-variant topological features along with time points.
Such patterns transformation corresponds with the transformation in the topological structure of time-evolving reconstructed networks.
Moreover, we quantify the transition points with respect to the topological structure by observing the sliding windows spanning two different developmental stages of \textit{Drosophila melanogaster}.
In each sliding window, we compute the kernel Fisher discriminant ratio~\cite{harchaoui:2009:kernel} 
for each time point from the three-dimensional 
persistence diagrams of the networks (see Appendix~\ref{sec:appx:fisher}).
The time point of the maximum ratio can be identified as the transition time point $t_c$ in each window~(Fig.~\ref{fig:kfdr}).
From the embryonic stage to the larval stage, 
we obtain the transition time points in the topological structure as $t_c=28$ and $t_c=31$,
relatively close to the experimentally known transition time point $t_1=31$.
Furthermore, from the larval stage to the pupal stage, 
and from the pupal stage to the adulthood stage, 
we obtain the transition time points in the topological structure as $t_c=41$ and $t_c=59$, respectively.
These points agree with the experimentally known transition time points $t_2=41$ and $t_3=59$.

\subsection{Considerations on the maximum value of the diffusion timescale}
We investigate the maximum timescale $\tmax$
to examine the length of the diffusion process must be explored.
Note that the timescale functions as a resolution parameter to unravel 
the multi-scale and hierarchical structure of a network.
A small timescale restricts random walkers in local interactions,
which produces many communities in the network.
In contrast, a large timescale leads to a substantial contribution of long walks 
and therefore yields a small number of communities
because random walkers tend to remain in these communities for a long time.
This resolution problem has been addressed in Refs.~\cite{delvenne:2010:stability,lambiotte:2014:modular}, 
in which the relevance of partitions as community structures is characterized over timescales.
Instead of characterizing the network structure at a fixed resolution,
the scale-variant topological features obtained with a sufficiently large $\tmax$ 
can contain information about the network at multiple resolutions.

Here, we study the method for determining $\tmax$ through the spectral decomposition of the normalized Laplacian of the network.
Denoting the eigenvalues of $\bLg$ by $\lambda_i$ in increasing order 
$0=\lambda_1\leq \lambda_2\leq \ldots \leq \lambda_n$,
the spectral decomposition of $\bLg$ is expressed as follows:
\begin{align}
    \bLg = \sum_{i=1}^n\lambda_i\bth^\top_i\bth_i,
\end{align}
where $\bth_i$ is the eigenvector associated with $\lambda_i$.
Therefore, the solution for Eq.~\eqref{eqn:master} can be written as follows:
\begin{align}\label{eqn:decomp}
    \bpg(\tau|i)=\sum_{i=1}^n\exp{(-\lambda_i\tau)}\bu_i\bth^\top_i\bth_i.
\end{align}
From Eq.~\eqref{eqn:decomp}, each eigenvalue $\lambda_i$ of $\bLg$ is associated with a decaying mode in the diffusion process
with the characteristic timescale $\tau=1/\lambda_i$.
Therefore, if there are large gaps between eigenvalues, for example if the $k_0$ smallest eigenvalues $\{\lambda_1=0,\ldots,\lambda_{k_0}\}$
are greatly separated from the remaining eigenvalues ($\lambda_{k_0} \ll \lambda_{k_0+1}=\lambda_{\textup{sep}}$), 
we can ignore the terms associated with $\{\lambda_{k_0+1},\ldots,\lambda_n\}$
in Eq.~\eqref{eqn:decomp} at $\tau$ satisfying $\tau\lambda_{\textup{sep}} \gg 1$. 
Thus, there is no significant change in the formation of clusters 
or loops in the mapped point cloud $P_{\cG}(\tau)$, 
nor in the formation of communities in the network at the $\tau$-scale.
Therefore, as a heuristic method,
if we consider $\tmax$ such that $\tmax\lambda_{\textup{sep}} \gg 1$, the structure of the network is well characterized via the scale-variant topological features.

\begin{figure}
		\includegraphics[width=8.5cm]{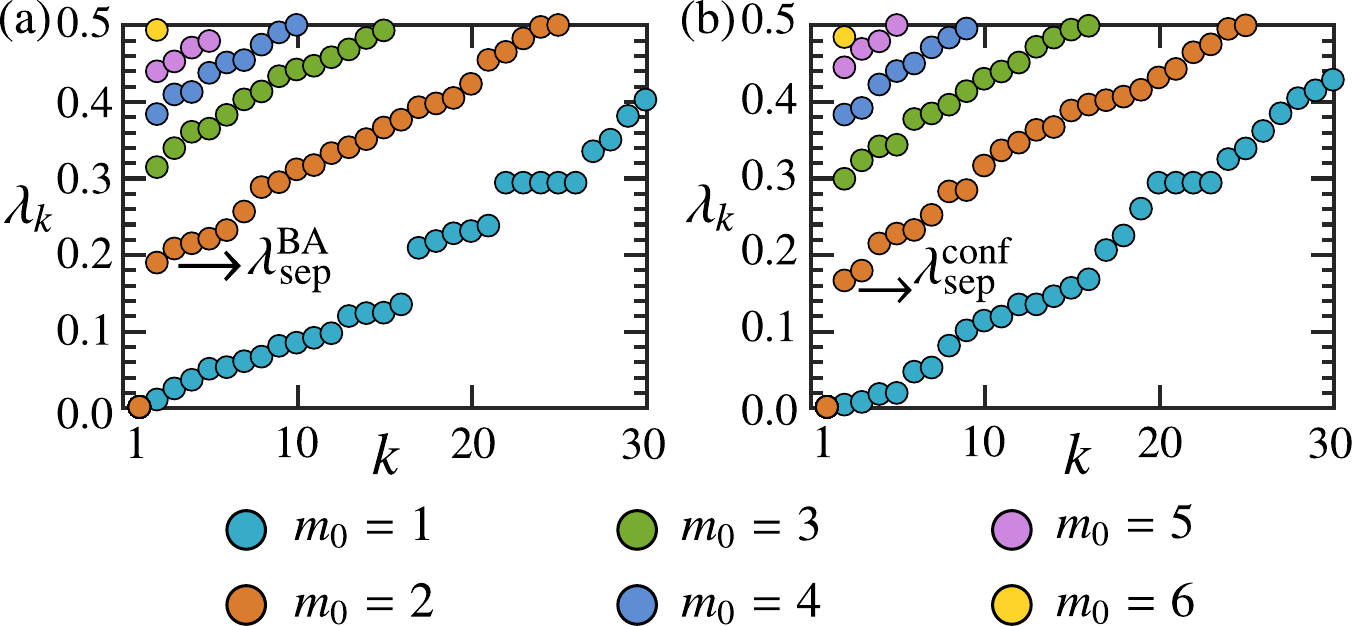}
		\protect\caption{Eigenvalues $\lambda_k$ such that $0\leq \lambda_k \leq 0.5$ for (a) Barab{\'a}si--Albert (BA) networks and (b) scale-free configuration networks. The colors of the points correspond to values of $m_0=1,2,3,4,5,6$ (number of initial nodes in the BA growth model).
		The value of $\lambda_{\textup{sep}}$ to effectively separate eigenvalues 
		is identified for each network. 
		The smallest $\lambda_{\textup{sep}}$ values for the BA networks and configuration networks are $\lambda^{\text{BA}}_{\textup{sep}}\approx 0.18$ and $\lambda^{\text{conf}}_{\textup{sep}}\approx 0.16$, respectively, as marked.
		\label{fig:specBA}}
\end{figure}

Here, we verify the above consideration by distinguishing the network of the Barab{\'a}si--Albert (BA) growth model~\cite{albert:1999:BA} with its configuration network.
The network is initialized with $m_0$ nodes and no edges. 
At each time step, each new node is added with no more than $m_0$ links to the existing nodes in the network. The probability that a new node is connected to an existing node is proportional to the degree of the existing node.
Note that both the BA networks and their configuration networks have a scale-free property with degree exponent 3.
We set the number of nodes to 128, vary the number of initial nodes $m_0=1,2, \ldots,50$, and generate 10 networks via this process for each value of $m_0$.
The 10 networks generated at each $m_0$ are split into two parts,
with five networks for training and the remaining five for testing.
Figure~\ref{fig:specBA} depicts eigenvalues $\lambda_k$ such that $0.0\leq \lambda_k \leq 0.5$ for (a) BA networks and (b) scale-free configuration networks. The colors of the points correspond to values of $m_0=1, 2, 3, 4, 5, 6$.
From Fig.~\ref{fig:specBA}, we can identify the value of $\lambda_{\textup{sep}}$ to separate the 
eigenvalues for each network.
The smallest $\lambda_{\textup{sep}}$ values for BA networks and configuration networks are $\lambda^{\text{BA}}_{\textup{sep}}\approx 0.18$ and $\lambda^{\text{conf}}_{\textup{sep}}\approx 0.16$, respectively.
Therefore, $\tmax$ should be set to $\tmax\lambda^{\text{BA}}_{\textup{sep}} \gg 1$ and $\tmax\lambda^{\text{conf}}_{\textup{sep}} \gg 1$,
for instance, $\tmax\lambda^{\text{BA}}_{\textup{sep}} > 10$ and $\tmax\lambda^{\text{conf}}_{\textup{sep}} > 10$, or $\tmax \geq 65$.

\begin{figure}
		\includegraphics[width=8.5cm]{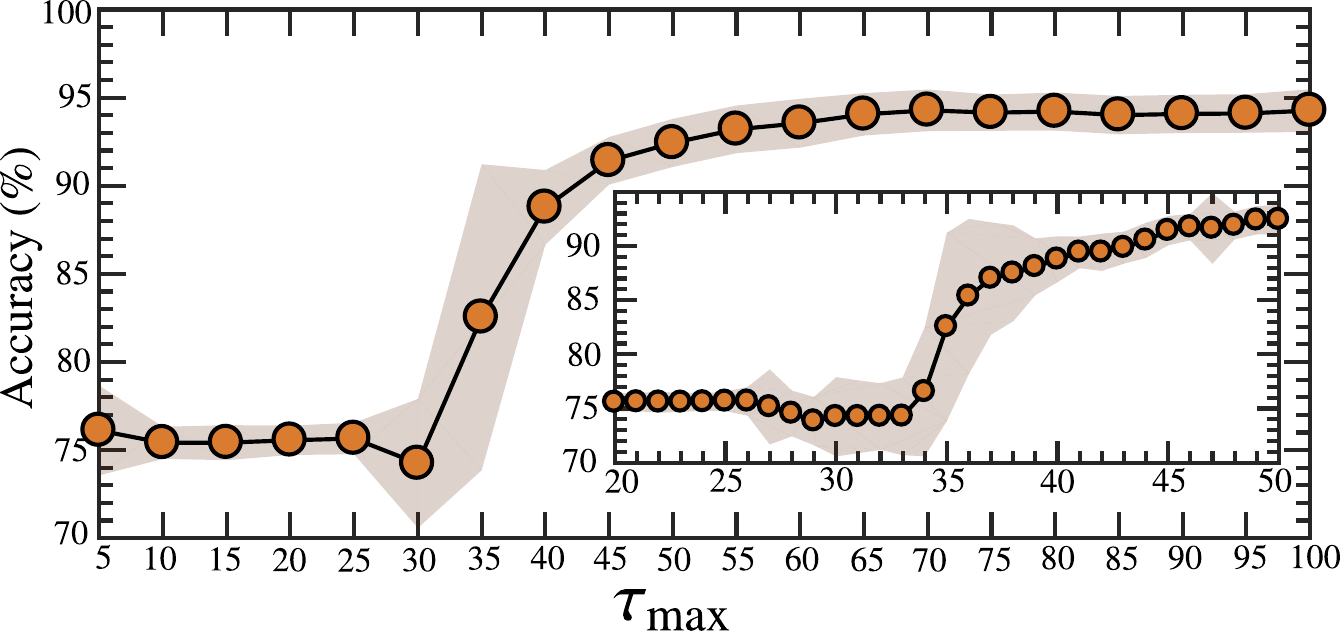}
		\protect\caption{Classification of networks generated from Barab{\'a}si--Albert models with their configuration networks using scale-variant topological features. The line depicts the average test accuracy over 100 random train-test splits at each value of $\tmax$. The shaded area indicates the confidence intervals of one standard deviation calculated using the same ensemble of runs. The inset highlights the transition in accuracy over $\tmax=20\text{--}50$.
		\label{fig:classifyBA}}
\end{figure}

For each $\tmax$ in $\{5, 10, 15, \ldots,95, 100\}$, 
we compute three-dimensional persistence diagrams of one-dimensional holes
with $\tau_1=1, \tau_2=2, \tau_3=3,\ldots, \tau_{K}=\tmax$.
The line in Fig.~\ref{fig:classifyBA} depicts the average test accuracy over 100 random train-test splits
at each value of $\tmax$.
The shaded area indicates the confidence intervals of one standard deviation calculated using the ensemble of runs.
In general, increasing $\tmax$ serves to increase classification accuracy because 
the diffusion process gathers more information about the network structure.
There is a transition in classification accuracy with large deviations when
$\tmax$ increases from 30 to 40.
To demonstrate this transition in more detail, 
we plot $\tmax=20, 21, \ldots, 49,50$ in the inset of Fig.~\ref{fig:classifyBA}.
For $\tmax < 30$, only microscale structures are considered in
the features, which has a small effect on the differences between networks.
The transition occurs when the mesoscale structures are considered.
For $\tmax > 40$, the deviation is reduced as the mesoscale structures are revealed.
For sufficiently large $\tmax$ ($\tmax \geq 65$), the method achieves high accuracy in the range of 94$\%$ to 94.5$\%$.
This observation agrees with the above-mentioned heuristic for determining $\tmax$.

\section{Concluding remarks and discussion}
Our study mainly aimed to represent the variation of topological scales, 
capture the nondyadic interactions, and provide robustness 
against noise in characterizing the structure of complex networks.
Here, we proposed a general framework for constructing the scale-variant 
topological features from the diffusion process exhibited in networks.
The scale-variant topological features do not directly correspond to the common statistical measures 
that are constructed from the dyadic interactions between nodes at a single fixed topological scale.
Rather, our features encode the information of both dyadic and nondyadic 
interactions in networks at variant topological scales.

In the networks classification, our features acted 
as strong factors to identify the networks.
Theoretically, we derived a strong mathematical guarantee for the robustness of 
these features with respect to the perturbations applied to the networks.
Through several experiments, we provided an empirical evidence 
for the effectiveness of these features in applications, 
such as classification of real-world networks and detection of transition points, 
with respect to the topological structure in time-evolving networks.
The results suggested that the observation of the topological features induced
from the network dynamics over variant scales
can characterize the structure and provide important insights for 
understanding the functionality of networks.

In our experiments, the scale-variant topological features were constructed from zero-dimensional and one-dimensional holes. 
In principle, we can compute the features from higher-dimensional holes that represent 
nondyadic interactions, involving a larger number of nodes in each interaction.
However, to investigate the features from high-dimensional holes,
the Vietoris--Rips filtration used in our study can consist of a large number of simplices.
More precisely, to consider $l$-dimensional holes, the Vietoris--Rips filtration 
has size $O(N^{l+2})$ of the number of simplices.
Herein, $N$ is the number of nodes in the network.
This observation shows the difficulty of using the features from $l$-dimensional holes ($l \geq 2$) 
for networks with a large number of nodes.
On the contrary and as demonstrated in this paper, we found it sufficient 
to use $l$-dimensional holes with $l=0, 1$ in practical applications.
Furthermore, one can replace the Vietoris--Rips filtration 
with the Witness filtration~\cite{silva:2004:witness} 
or the approximation of the Vietoris--Rips filtration~\cite{sheehy:2013:linear} 
for more efficient computations.
We employed recent algorithmic improvements
to efficiently compute the persistent diagrams with the core implementation referenced from Ripser libary~\cite{bauer:2017:ripser}.

Another point to discuss is the selection of the maximum diffusion timescale.
In our experiments, our method was only tested for small and medium-sized networks with 
less than 5,000 nodes per network.
For larger networks, a longer diffusion timescale must be explored.
This consideration will increase the computational cost of computing persistence diagrams and the kernel.
However, this limitation can be mitigated by increasing the sampling interval to take discrete values of the timescale
while keeping the maximum timescale sufficiently large.
It is also possible to study the process of taking values of timescales based on
the spectral decomposition of the normalized Laplacian of the network.

Our study is motivated by Ref.~\cite{domenico:2017:diffusion}, 
in which the diffusion geometry from the diffusion process
is used to reveal functional clusters in a network.
Based on random walk dynamics, the diffusion distance between a pair of 
nodes in a network is defined and averaged in a range of 
timescales to model the underlying geometry of the network.
However, the variation in network structure over the diffusion timescale is not discussed.
In Ref.~\cite{lambiotte:2014:modular}, a random walk process corresponding to the natural dynamics of a system focuses on recovering dynamically meaningful communities in the network.
Our approach involves a diffusion process similar to that mentioned in Refs.~\cite{lambiotte:2014:modular,domenico:2017:diffusion}; 
however, it mainly focuses on the systematic representation of networks via topological data analysis by tracking the variation of the topological structures 
along timescales of the diffusion process.

In general, our study presented a unified analysis of complex networks.
This study paves several opportunities for designing effective algorithms in network science,
such as an investigation of more complicated network structures.
For instance, we can employ our framework to study different aspects of multiplex networks, 
or to study the structural reducibility of a multilayer network while preserving its dynamics and function.

\section{Acknowledgments}
This work was supported by Ministry of Education, Culture, Sports, Science and Technology (MEXT) KAKENHI Grants No. JP16K00325 and No. JP19K12153.

\appendix

\section{Construction of Vietoris--Rips filtration of a network\label{sec:appx:filt}}
\begin{figure*}
	\begin{center}
		\includegraphics[width=16cm]{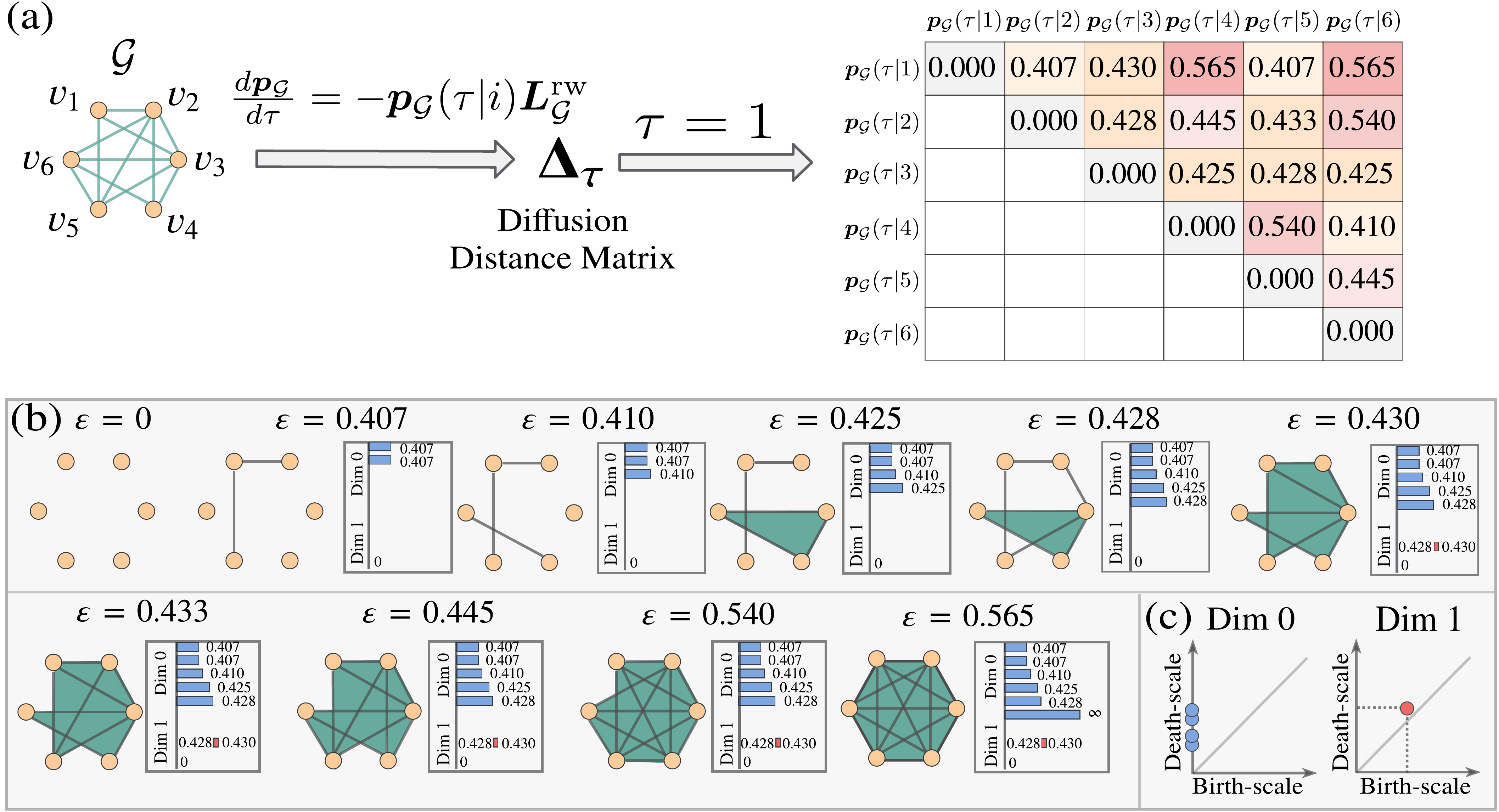}
		\protect\caption{
		An exemplary of Vietoris--Rips filtration constructed from a complex network $\cG$ at a specific timescale $\tau$ ($\tau=1$ in this example).
		We map nodes $v_1, \ldots, v_N$ of the network $\cG$ to a point cloud of $N$ points $\bpg(\tau|1), \ldots, \bpg(\tau|N)$ through a diffusion dynamics described by the random walk Laplacian $\bLg$.
		(a) Diffusion distance matrix $\bDelta_\tau$ of size $N\times N$, whose element $\Delta_{ij}$ is the Euclidean distance between points $\bpg(\tau|i)$ and $\bpg(\tau|j)$.
		(b) A complex is built over a set of points if the pairwise distances between them are less than or equal to a threshold parameter $\varepsilon$.
		If $\varepsilon=0$, we have the discrete points. 
		As $\varepsilon$ takes the increasing sequence values of diffusion distance $\Delta_{ij}$, the hole patterns such as connected components (zero-dimensional hole) or loops (one-dimensional hole) can appear or disappear over this filtration.
		The lifetime of these hole patterns are described as blue bars (for zero-dimensional holes) and red bars (for one-dimensional holes).
		These bars begin at the values of $\varepsilon$ when the holes appear, then end at values when the holes disappear. 
		(c) The corresponding persistence diagrams for zero-dimensional and one-dimensional holes.
		The birth-scale and the death-scale are represented for the values of $\varepsilon$ at the emergence and the disappearance of the holes.
		\label{fig:scale-demo}}
	\end{center}
\end{figure*}

We define and describe in Fig.~\ref{fig:scale-demo} 
the process of extracting topological features of a complex network at each specific timescale $\tau$. 
At each $\tau$, we calculate the diffusion distance matrix $\bDelta_\tau$ of size $N\times N$, 
whose element $\Delta_{ij}$ is the Euclidean distance between points $\bpg(\tau|i)$ and $\bpg(\tau|j)$ [Fig.~\ref{fig:scale-demo}(a)].
If $\varepsilon = 0$, the nodes of the network can be considered discrete points.
As we increase $\varepsilon$, new pairwise connections and simplices may appear 
when $\varepsilon$ meets each value of $\Delta_{ij}$.
We obtain a filtration as a sequence of embedded simplicial complexes.
Hole patterns such as connected components (zero-dimensional holes) or loops (one-dimensional holes) 
can appear or disappear over this filtration.
For instance, in   Fig.~\ref{fig:scale-demo}(b), at $\varepsilon=0$, 
we have six separated nodes considered as six separated connected components, 
but at $\varepsilon=0.407$, three nodes are connected with each other; 
thus, two connected components disappear at this scale.
We can describe these patterns as two blue bars started at scale $0$ and ended at scale $0.407$.
The same explanation with the red bar started at scale $0.428$ and ended at scale $0.430$, 
which represents the emergence of loop pattern ($v_1\rightarrow v_2 \rightarrow v_3 \rightarrow v_5 \rightarrow v_1$) 
at $\varepsilon=0.428$ and the disappearance at $\varepsilon=0.430$.
Figure~\ref{fig:scale-demo}(c) illustrates the corresponding persistence diagrams 
for zero-dimensional holes and one-dimensional holes, where the birth-scale and the death-scale 
are represented for the values of $\varepsilon$ at the emergence and the disappearance of the holes.

\section{Proof of the stability of the scale-variant topological features\label{sec:appx:stab}}
We prove the result in Eq.~\eqref{eqn:bottleneck:stablility}.
First, we introduce the concept of the bottleneck distance between two two-dimensional diagrams. 
Let $X$ and $Y$ be finite sets of points embedded in the Euclidean space $\mathbb{R}^n$. 
Denote their $two$-dimensional persistence diagrams for $l$-dimensional holes as $\dgX$ and $\dgY$, 
respectively.  We consider all matchings, $\gamma$, such that a point on one diagram is matched to 
a point on the other diagram or to its projection on the line $b=d$ in two-dimensional space. 
The bottleneck distance $\dBot$ between $\dgX$ and $\dgY$ is defined as 
the infimum of the longest matched infinity-norm distance over all matchings, $\gamma$:
	\begin{equation}\label{eqn:bottleneck2:def}
	\dBot(\dgX,\dgY) = \inf_{\gamma} \max_{(\bqa,\bqb) \in \gamma} \Vert \bqa- \bqb \Vert_{\infty}.
	\end{equation} 
Here, $\Vert \bqa- \bqb \Vert_{\infty} = \max\left(|b_1-b_2|, |d_1-d_2|\right)$ for which $\bqa=(b_1, d_1)$ and $\bqb=(b_2, d_2)$.

The bottleneck distance between the two-dimensional persistence diagrams 
	satisfies the following inequality~\cite{chazal:2014:stab}: 
	\begin{equation}\label{eqn:stab:2D}
		\dBot(\dgX, \dgY) \leq 2\dHau(X, Y),
	\end{equation}
	where $\dHau(X, Y)$ is the Hausdorff distance given as 
	\begin{equation}
		 \dHau(X, Y) = \max \left\{ \max_{\bx\in X} \min_{\by \in Y} d(\bx, \by),  \max_{\by \in Y} \min_{\bx \in X}d(\bx, \by)\right\}.
	\end{equation}
	Here, $d(\bx, \by)$ is the Euclidean distance between two points $\bx, \by$ in $\mathbb{R}^n$.

Given two three-dimensional persistence diagrams as $E$ and $F$, 
consider all matchings $\psi$ such that a point on one diagram is matched 
to a point on the other diagram or to its projection on the plane $b=d$.
For each pair $(\bqa, \bqb) \in \psi$ for which $\bqa=\left(b_1,d_1,\tau_1\right)$ and $\bqb=\left(b_2,d_2,\tau_2\right)$, 
we define \textit{the relative infinity-norm distance} 
between $\bqa$ and $\bqb$ as $d^{(\infty)}_{\xi}(\bqa, \bqb)=\max\left(|b_1-b_2|, |d_1-d_2|, \xi|\tau_1-\tau_2|\right)$,
where $\xi$ is a positive rescaling coefficient introduced to adjust 
the scale difference between the point-wise distance and time.
The bottleneck distance, $\dBBxi(E, F)$, is defined as the infimum of 
the longest matched relative infinity-norm distance over all matchings $\psi$:
\begin{equation}\label{eqn:bottleneck:def}
\dBBxi(E, F) = \inf_{\psi} \max_{(\bqa,\bqb) \in \psi} d^{(\infty)}_{\xi}(\bqa, \bqb).
\end{equation}

For each $\tau \in \mathcal{T}$ and two networks $\mathcal{G}, \mathcal{H}$ with the same number $N$ of nodes, we first prove the following inequality:
\begin{align}
    \dBot(\dgtG, \dgtH) \leq 2\tau \Vert \bLg - \bLh \Vert_2
    \label{eqn:ineq:tau}.
\end{align}
Here, two two-dimensional persistence diagrams $D^{(2)}_{l,\tau}(\mathcal{G})$ 
and $D^{(2)}_{l,\tau}(\mathcal{H})$ are calculated for $l$-dimensional holes 
from two point clouds $\bPGtau=\{\bpg(\tau|1), \ldots,\bpg(\tau|N)\}$ 
and $\bPHtau=\{\bph(\tau|1), \ldots,\bph(\tau|N)\}$, respectively.

Since $\dgtG = \cS_{(l)}(\bPGtau)$ and $\dgtH = \cS_{(l)}(\bPHtau)$,
we apply Eq.~\eqref{eqn:stab:2D} to have
\begin{align}
    \dBot(\dgtG, \dgtH) &=  \dBot (\cS_{(l)}(\bPGtau), \cS_{(l)}(\bPHtau)) \nonumber\\
    &\leq 2\dHau(\bPGtau, \bPHtau).\label{eqn:botneck:tau}
\end{align}

From the definition of the Hausdorff distance, we have
\begin{widetext}
\begin{align}
    \dHau(\bPGtau, \bPHtau) &= \max \left\{ \max_{i} \min_{j} d(\bpg(\tau|i), \bph(\tau|j)),  \max_{j} \min_{i} d(\bpg(\tau|i), \bph(\tau|j)))\right\} \\
     &\leq \max \left\{ \max_{i} d(\bpg(\tau|i), \bph(\tau|i)),  \max_{j} d(\bpg(\tau|j), \bph(\tau|j)) \right\} = \max_{i} d(\bpg(\tau|i), \bph(\tau|i)).\label{eqn:ineq:hausdorff}
\end{align}
\end{widetext}

Since $\bpg(\tau|i)=\bu_i\exp(-\tau\bLg)$ and $\bph(\tau|i)=\bu_i\exp(-\tau\bLh)$, we have
\begin{align}
    d(\bpg(\tau|i), \bph(\tau|i)) &= \Vert \bu_i (e^{-\tau\bLg} - e^{-\tau\bLh}) \Vert_2 \\
    &\leq \Vert \bu_i \Vert_2\Vert e^{-\tau\bLg} - e^{-\tau\bLh} \Vert_2 \\
    &= \Vert e^{-\tau\bLg} - e^{-\tau\bLh} \Vert_2.\label{eqn:ineq:disttau1}
\end{align}
We write the difference of the matrix exponential in terms of an integral~\cite{golub:2012:matrix},
\begin{align}
    \Vert e^{-\tau\bLg} - e^{-\tau\bLh} \Vert_2 &= \bigg\Vert \int_{0}^{\tau}e^{-\bLg(\tau-t)}\bE e^{-\bLh t}dt \bigg\Vert_2\\
    &\leq \int_{0}^{\tau} \Vert e^{-\bLg(\tau-t)}\bE e^{-\bLh t} \Vert_2 dt \\
    &\leq \Vert \bE \Vert_2 \int_{0}^{\tau} \Vert e^{-\bLg(\tau-t)}\Vert_2 \Vert e^{-\bLh t} \Vert_2 dt,\label{eqn:ineq:disttau2}
\end{align}
where $\boldsymbol{E}=\bLg - \bLh$.
We know that $-\bLg$ is a negative semi-definite matrix with the largest eigenvalue equal to 0.
It implies that the largest eigenvalue of $e^{-\bLg(\tau-t)}$ is equal to 1, and hence  $\Vert e^{-\bLg(\tau-t)}\Vert_2 = 1$. We obtain the same result with   $-\bLh$, i.e., $\Vert e^{-\bLh t} \Vert_2 = 1$. Then, from Eq.~\eqref{eqn:ineq:disttau1} and Eq.~\eqref{eqn:ineq:disttau2}, we have
\begin{align}
    d(\bpg(\tau|i), \bph(\tau|i)) &\leq \Vert \bE \Vert_2 \int_{0}^{\tau} 1dt \\
    &= \tau\Vert \bE \Vert_2 = \tau\Vert \bLg - \bLh \Vert_2.\label{eqn:ineq:disttau3}
\end{align}
From Eq.~\eqref{eqn:botneck:tau}, Eq.~\eqref{eqn:ineq:hausdorff} and Eq.~\eqref{eqn:ineq:disttau3}, we have the result in Eq.~\eqref{eqn:ineq:tau}.

Let $\Gamma_{\tau}$ be the set of matchings defined in Eq.~\eqref{eqn:bottleneck2:def} between two two-dimensional persistence diagrams $\dgtG$ and $\dgtH$. 
For each collection $\Lambda=\{\gamma_1, \gamma_2, \ldots, \gamma_K  \mid \gamma_i \in \Gamma_{\tau_i}, i=1,2,\ldots, K\}$, 
we construct the matching $\psi$ between two three-dimensional persistence diagrams $\dgG$ and $\dgH$, 
such that, for each $(\bqa, \bqb) \in \psi$, then $\bqa=(b_1, d_1, \tau)$, $\bqb=(b_2, d_2, \tau)$,
	and $(\bqa_{\gamma}, \bqb_{\gamma}) \in \gamma$, where $\bqa_{\gamma}=(b_1, d_1)$, $\bqb_{\gamma}=(b_2, d_2)$ 
	and $\gamma \in \Lambda \cap \Gamma_{\tau}$.
Let $\Gamma$ be a set of all matchings $\psi$ constructed this way. 
From the definition of the bottleneck distance, we have the following inequality:
	 
	\begin{equation} \label{eqn:bttdis:inf}
		\dBBxi(\dgG, \dgH) \leq \inf_{\psi \in \Gamma} \max_{(\bqa, \bqb) \in \psi} d^{(\infty)}_{\xi}(\bqa, \bqb).
	\end{equation}
    
For $(\bqa, \bqb) \in \psi$, we have
	\begin{align}
			d^{(\infty)}_{\xi}(\bqa, \bqb) &= \max\{|b_1-b_2|, |d_1-d_2|, \xi|\tau-\tau|\}\\
			 &= \max\{|b_1-b_2|, |d_1-d_2|\}\\
			 &= \Vert \bqa_{\gamma} - \bqb _{\gamma}\Vert_{\infty},
	\end{align}
and Eq.~\eqref{eqn:bttdis:inf} becomes
	\begin{align} 
		\dBBxi(\dgG,\dgH) & \leq \max_{\tau \in \mathcal{T}} \inf_{\gamma \in \Gamma_{\tau}} \max_{(\bqa_{\gamma}, \bqb _{\gamma}) \in \gamma} \Vert \bqa_{\gamma} - \bqb _{\gamma}\Vert_{\infty}\\
	 &= \max_{\tau \in \mathcal{T}} \dBot(\dgtG, \dgtH).\label{eqn:bttdis:2to3}
	\end{align}
From Eq.~\eqref{eqn:ineq:tau} and Eq.~\eqref{eqn:bttdis:2to3}, we obtain Eq.~\eqref{eqn:bottleneck:stablility} in the main text, which is the stability property of our scale-variant features.

\section{Selecting parameters for the kernel\label{sec:appx:param}}
In the kernel $\cK_{\sigma, \xi}$ defined in Eq.~\eqref{eqn:kernel},
we set the rescale coefficient to $\xi=\sigma$ and present here a heuristic method to select the bandwidth $\sigma$.
Given the kernel values calculated from the three-dimensional persistence diagrams $D_{(l),1}^{(3)}, D_{(l),2}^{(3)}, \ldots, D_{(l),M}^{(3)}$ of $l$-dimensional holes,
we denote $\sigma^2_s = \text{median}\{(b_i-b_j)^2+(d_i-d_j)^2 \mid (b_i, d_i, \tau_i), (b_j, d_j, \tau_j) \in D_{(l), s}^{(3)}\}$ with $s=1,2,\ldots,M$.
	We set $\sigma$ as $\sigma^2 = \dfrac{1}{2}\text{median}\{\sigma^2_s \mid s = 1,\ldots,M\}$ such that $2\sigma^2$ takes values close to many $(b_i-b_j)^2+(d_i-d_j)^2$ values.
	
\section{Kernel Fisher discriminant ratio\label{sec:appx:fisher}} 
Consider a collection of three-dimensional diagrams  $\mathcal{D}_{(l)}=\{D_{(l),1}^{(3)},D_{(l),2}^{(3)},\ldots,D_{(l),M}^{(3)}\}$ of $l$-dimensional holes.
Since $\cK_{\sigma, \xi}$ is a positive-definite kernel on $\mathcal{D}_{(l)}$~\cite{tran:2018:variant}, there exists a Hilbert space $H_b$ and a mapping $\Phi: \mathcal{D}_{(l)}\longrightarrow H_b$ such that for $E \in \mathcal{D}_{(l)}$, $\Phi$ maps $E$ to a function  $\Phi_E\in H_b$ that satisfies
\begin{align}
    \forall E, F \in \mathcal{D}_{(l)}, \cK_{\sigma, \xi}(E, F) = \langle \Phi_E, \Phi_F \rangle_{H_b}.
\end{align}
Here, $H_b$ is a real inner product space of function $f: \mathcal{D}_{(l)}\longrightarrow \mathbb{R}$, and thus, is a complete metric space with respect to the
distance induced by the inner product $\langle \cdot, \cdot \rangle_{H_b}$.

Given an index $s>1$, the kernel Fisher discriminant ratio $\kfdr_{M,s}(\mathcal{D}_{(l)})$ is a statistical quantity to measure 
the dissimilarity between two classes assumptively defined by two sets of diagrams 
having index before and from $s$~\cite{harchaoui:2009:kernel}.
The corresponding empirical mean and covariance functions in $H_b$ associated with the data in 
$\mathcal{D}_{(l)}$ having index before and from $s$ are defined as
\begin{align}
    \hat{\mu}_1 &= \dfrac{1}{s-1}\sum_{i=1}^{s-1}\Phi_{D_{(l),i}^{(3)}}, \\
    \hat{\Sigma}_1 &= \dfrac{1}{s-1}\sum_{i=1}^{s-1}\left\{ \Phi_{D_{(l),i}^{(3)}} - \hat{\mu}_1 \right \}  \otimes \left\{ \Phi_{D_{(l),i}^{(3)}} - \hat{\mu}_1 \right \},\\
    \hat{\mu}_2 &= \dfrac{1}{M-s+1}\sum_{i=s}^{M}\Phi_{D_{(l),i}^{(3)}}, \\
    \hat{\Sigma}_2 &= \dfrac{1}{M-s+1}\sum_{i=s}^{M}\left\{ \Phi_{D_{(l),i}^{(3)}} - \hat{\mu}_2 \right \}  \otimes \left\{ \Phi_{D_{(l),i}^{(3)}} - \hat{\mu}_2 \right \}.
\end{align}
Here, $f\otimes g$ for two functions $f, g \in H_b$ is defined for all functions $h \in H_b$ as $(f\otimes g)h = \left < g, h\right >_{H_b}f$. 

The kernel Fisher discriminant ratio $\kfdr_{M,s}(\mathcal{D}_{(l)})$ is defined as
\begin{align}
    &\kfdr_{M,s}(\mathcal{D}_{(l)}) \nonumber\\
    &=\dfrac{(s-1)(M-s+1)}{M}\left<\hat{\mu}_2 - \hat{\mu}_1, (\hat{\Sigma} + \eta \bI)^{-1}(\hat{\mu}_2 - \hat{\mu}_1)\right>_{H_b},
\end{align}
where $\eta$ is a regularization parameter and 
    $\hat{\Sigma} = \dfrac{s-1}{M}\hat{\Sigma}_1 + \dfrac{M-s+1}{M}\hat{\Sigma}_2$.
Here, the index $s$ achieving the maximum of $\kfdr_{M,s}(\mathcal{D}_{(l)})$ 
corresponds to the estimated transition point.

We set $\eta=10^{-1}, 10^{-1}, 10^{-5}$ in the experiments of the Girvan--Newman (GN) network, Lancichinetti--Fortunato--Radicchi (LFR) network and \textit{Drosophila melanogaster} network, respectively.

\section{Common measures for a network\label{sec:appx:common}}
For each network, we calculate the following 18 common measures: the density (the ratio of the existing to the possible edges), the transitivity~\cite{costa:2007:characterization} (the proportion of triangles), the diameter (the maximum eccentricity), 
the radius (the minimum eccentricity), the degree assortativity coefficient~\cite{newman:2002:assortative}, 
the global efficiency~\cite{latora:2001:efficient}, the number of connected parts, 
the average clustering coefficient,
the average number of triangles that include a node as a vertex, 
the average local efficiency~\cite{latora:2001:efficient}, the average edge betweenness
centrality~\cite{brandes:2008:variants}, the average node betweenness centrality~\cite{freeman:1978:centrality}, 
the average node closeness centrality~\cite{freeman:1978:centrality}, the average eccentricity,
the average shortest paths, the average degree centrality~\cite{freeman:1978:centrality},
the maximum modularity which is obtained by Louvain heuristic~\cite{blondel:2008:unfold,girvan:2004:benchmark}, 
and the average of global mean first-passage times of random walks on the network~\cite{tejedor:2009:passage}.
We normalize the measures in the range of  $[0, 1]$ using the min-max normalization (i.e. $f_{*}=(f_{*}-f_{\text{min}})/(f_{\text{max}}-f_{\text{min}})$, 
where $f_{\text{min}}, f_{\text{max}}$ are the minimum and the maximum values of a measure in the data).

\section{Graph kernel methods\label{sec:appx:graphker}} 
We describe the graph kernel methods used in the main text.
The implementations of these graph kernels can be found in Ref.~\cite{sugiyama:2017:graphkernels}.

\subsection{Random walk kernels} 
The random walk graph kernels measure the similarity between a pair of graphs
based on the number of equal-length walks in two graphs.
Given two unlabeled graphs $\mathcal{G}$ and $\mathcal{G}'$ with their vertex and edge sets as $(\mathcal{V}, \mathcal{E})$ and $(\mathcal{V}', \mathcal{E}')$, respectively, the direct product graph $\mathcal{G}_{\times}=(\mathcal{V}_{\times}, \mathcal{E}_{\times})$ of $\mathcal{G}$ and $\mathcal{G}'$ 
is a graph with the node set $\mathcal{V}_{\times}=\{(v, v') \mid v\in \mathcal{V}, v' \in \mathcal{V}' \}$ 
and the edge set $\mathcal{E}_{\times}=\{ ((v_{a}, v'_{a}), (v_{b}, v'_{b})) \mid (v_{a}, v_{b}) \in \mathcal{E}, (v'_{a}, v'_{b}) \in \mathcal{E}'\}$.

The KStepRW kernel is the $k$-step random walk kernel $\mathcal{K}_{\times}^{k}$ defined as
\begin{align}
    \mathcal{K}_{\times}^k(\mathcal{G}, \mathcal{G}')=\sum_{i,j=1}^{|\mathcal{V}_{\times}|}\sum_{m=0}^k\left[ \lambda_m\bW_{\times}^m \right]_{ij},
\end{align}
where $\bW_{\times}$ is a weight matrix of $\mathcal{G}_{\times}$ and $\lambda_0, \ldots, \lambda_k$ is a sequence of positive, real-valued weights.
In our experiments, we set $k=2$ and $\lambda_0=\lambda_1=\lambda_2=1.0$.

GeometricRW kernel is a specific case of the $k$-step random walk kernel, when $k$ goes to infinity 
and the weights are the geometric series, i.e., $\lambda_m=\lambda^m$ ($\lambda=0.05$ in our experiments). 
The GeometricRW kernel is defined as
\begin{align}
    \mathcal{K}_{\text{GR}}(\mathcal{G}, \mathcal{G}')=\sum_{i,j=1}^{|\mathcal{V}_{\times}|}\sum_{m=0}^\infty\left[ \lambda^m\bW_{\times}^m \right]_{ij}=\sum_{i,j=1}^{|V_{\times}|}\left[ (\bI - \lambda \bW_{\times})^{-1}\right],
\end{align}
where $\bI$ is an identity matrix of size $|\mathcal{V}_{\times}| \times |\mathcal{V}_{\times}|$.

ExponentialRW kernel is a specific case of the $k$-step random walk kernel, when $k$ goes to infinity and the weights are the exponential series, i.e., $\lambda_m=\frac{\beta^m}{m!}$ ($\beta=0.1$ in our experiments).
The ExponentialRW kernel is defined as
\begin{align}
    \mathcal{K}_{\text{EX}}(\mathcal{G}, \mathcal{G}')=\sum_{i,j=1}^{|V_{\times}|}\sum_{m=0}^\infty\left[ \dfrac{(\beta\bW_{\times})^m}{m!} \right]_{ij}=\sum_{i,j=1}^{|V_{\times}|}\left[ e^{\beta\bW_{\times}}\right]_{ij}.
\end{align}

\subsection{ShortestPath kernel} 
ShortestPath kernel compares all pairs of the shortest path lengths from $\mathcal{G}$ and $\mathcal{G}'$ defined as
\begin{equation}
    \mathcal{K}_{\text{SP}}(\mathcal{G}, \mathcal{G}')=\sum_{v_i,v_j \in \mathcal{G}}\sum_{v'_k,v'_l \in \mathcal{G}'}\delta(d(v_i, v_j), d(v'_k, v'_l)),
\end{equation}
where $d(v_i, v_j)$ and $d(v'_k, v'_l)$ are the lengths of the shortest path between nodes $v_i$ and $v_j$ in $\mathcal{G}$, and the shortest path between nodes $v'_k$ and $v'_l$ in $\mathcal{G}'$, respectively.
Here, $\delta(x,y)=1$ if $x=y$, and $0$ if $x \neq y$.

\subsection{Graphlet kernel} 
A size-$k$ graphlet is an induced and non-isomorphic sub-graph of size $k$. 
Let $\mathcal{S}_k=\{G_1,\ldots,G_{N_k}\}$ be a set of size-$k$ graphlets, where $N_k$ denotes the number of unique graphlets of size $k$.
For an unlabeled graph $\mathcal{G}$ (the graph does not contain attributes for nodes), we define a vector $\boldsymbol{f}_{\mathcal{G}}$ of length $N_k$ such that the $i^{th}$ component of $\boldsymbol{f}_{\mathcal{G}}$ denotes the frequency of graphlet $G_i$ appearing as a subgraph of $\mathcal{G}$.
Given two unlabeled graphs $\mathcal{G}$ and $\mathcal{G}'$, the graphlet kernel is defined as
\begin{align}
   \mathcal{K}_{\text{GK}}(\mathcal{G}, \mathcal{G}') = \langle\boldsymbol{f}_{\mathcal{G}},\boldsymbol{f}_{\mathcal{G}'}\rangle,
\end{align}
where $\langle\cdot,\cdot\rangle$ represents the Euclidean dot product.
We set $k=4$ for MUTAG, BZR, DHFR and FRANKENSTEIN datasets and $k=3$ for the other datasets.

\subsection{Weisfeiler--Lehman kernel} 
Weisfeiler--Lehman kernel decomposes a graph into its subtree patterns and compares these patterns in two graphs.
For an unlabeled graph $\mathcal{G}$, all vertexes $v$ of $\mathcal{G}$ are initialized with label $\varphi(v)=0$.
We iterate over each vertex $v$ and its neighbour to create a multiset label as $\varphi^{(i)}(v)$ such that $\varphi^{(1)}(v) = \varphi(v)$, and $\varphi^{(i)}$ with $i>1$ is defined as $\varphi^{(i)}(v) = (\varphi^{(i)}(v), \text{Q}^{(i-1)}_v)$, where $\text{Q}^{(i-1)}_v$ is the sorted labels of $v$'s neighbours.
To measure the similarity between graphs, we count the co-occurrences of the labels in both graphs for $h$ iterations with the kernel defined as
\begin{align}
    \mathcal{K}_{\text{WL}}(\mathcal{G}, \mathcal{G}')=\langle\boldsymbol{1}_{\mathcal{G}},\boldsymbol{1}_{\mathcal{G}'}\rangle.
\end{align}
Here, $\boldsymbol{1}_{\mathcal{G}}$ is the vector concatenation of $h$ vertex label histograms  $\boldsymbol{1}^{(1)}_{\mathcal{G}}, \ldots, \boldsymbol{1}^{(h)}_{\mathcal{G}}$ in $h$ iterations.
We set $h=5$ in our experiments.

\providecommand{\noopsort}[1]{}\providecommand{\singleletter}[1]{#1}%
\end{document}